\documentclass[8pt]{article}
\usepackage{graphicx}
\usepackage{natbib}

\bibpunct{(}{)}{;}{a}{,}{,}

\newcommand{\be}{\begin{equation}}
\newcommand{\ee}{\end{equation}}

\begin{document}

\title{The role of dilation and confining stresses in shear thickening of dense suspensions}
\author{Eric Brown and Heinrich M. Jaeger\\
James Franck Institute, The University of Chicago, Chicago, IL 60637
}

\maketitle

\begin{abstract}

Many densely packed suspensions and colloids exhibit a behavior known as Discontinuous Shear Thickening in which the shear stress jumps dramatically and reversibly as the shear rate is increased.  We performed rheometry and video microscopy measurements on a variety of suspensions to determine the mechanism for this behavior.  We distinguish Discontinuous Shear Thickening from inertial effects by showing that the latter are characterized by a Reynolds number but are only found for lower packing fractions and higher shear rates than the former.  Shear profiles and normal stress measurements indicate that, in the shear thickening regime, stresses are transmitted through frictional rather than viscous interactions, and come to the surprising conclusion that for concentrated suspensions such as cornstarch in water which exhibit the phenomenon of Discontinuous Shear Thickening, the local constitutive relation between stress and shear rate is not necessarily shear thickening.  If the suspended particles are heavy enough to settle we find the onset stress of shear thickening $\tau_{min}$ corresponds to a hydrostatic pressure from the weight of the particle packing where neighboring particles begin to shear relative to each other.  Above $\tau_{min}$, dilation is seen to cause particles to penetrate the liquid-air interface of the sheared sample.  The upper stress boundary $\tau_{max}$ of the shear thickening regime is shown to roughly match the ratio of surface tension divided by a radius of curvature on the order of the particle size.   These results suggest a new model in which the increased dissipation in the shear thickening regime comes from frictional stresses that emerge as dilation is frustrated by a confining stress from surface tension at the liquid-air interface.   We generalize this shear thickening mechanism to other sources of a confining stress by showing that, when instead the suspensions are confined by solid walls and have no liquid-air interface, $\tau_{max}$ is set by the stiffness of the most compliant boundary which frustrates dilation.  All of this rheology can be described by a non-local constitutive relation in which the local relation between stress and shear rate is shear thinning, but where the stress increase comes from a normal stress term which depends on the global dilation. 

 \end{abstract}



\section{Introduction}

\small

Shear thickening is a category of non-Newtonian fluid behavior in which the viscosity $\eta=\tau/\dot\gamma$ increases as a function of shear rate $\dot\gamma$ or shear stress $\tau$ over some parameter range.  A particularly dramatic manifestation characterized by a sharp jump in stress with increasing shear rate, often called Discontinuous Shear Thickening \citep{MW58, Ho72, Ba89, MW01a}, occurs in many densely packed suspensions and colloids such as cornstarch in water.    These suspensions feel like a thin liquid at low stresses, but become very thick and can even crack like a solid at higher stresses, and become thin again when the stress is removed.  Such materials are of practical interest for their properties as dampeners and shock absorbers \citep{LWW03, SWB03, Lord}.  While some milder types of shear thickening can be explained as viscous \citep{BB85, WB09, CMIC11} or inertial \citep{Ba54} effects, prior approaches have not been very successful at describing the dramatic effects of Discontinuous Shear Thickening.  Our goal with this paper is to use observations to develop and test a model for Discontinuous Shear Thickening.  This problem can be broken down into several questions: under what conditions will a suspension exhibit Discontinuous Shear Thickening?,  what are the scaling laws that determine the parameter range of shear thickening?, and what is the form of the constitutive law? We will use a wide variety of rheometry and video microscopy measurements to answer the above questions.  Our approach differs from previous work in that we consider a granular point of view in addition to the traditional hydrodynamic approach by investigating the source of observed stresses during Discontinuous Shear Thickening, rather than focusing only on the viscous contribution to the local constitutive relation.


The remainder of this paper is organized as follows.  In Sec.~\ref{sec:background} we review the literature on shear thickening to characterize  Discontinuous Shear Thickening and summarize existing models.  In Sec.~\ref{sec:methods} we describe the rheometry techniques and suspensions used in our experiments.  In Sec.~\ref{sec:inertial} we show viscosity curves for shear thickening suspensions at different packing fractions and liquid viscosities.  With this we can test inertial hydrodynamic scalings which we find to apply in a different parameter regime than Discontinuous Shear Thickening.  In Sec.~\ref{sec:onsetstress} we show measurements of the onset stress $\tau_{min}$ of shear thickening for different particle sizes and liquid densities and show that for settling suspensions, the onset of shear thickening occurs when the shear stress becomes large enough to initiate relative shear between particles, balancing against a gravitational pressure.  In Sec.~\ref{sec:shearprofile} we show shear profile measurements with levels of density matching and demonstrate that the large stress jump that characterizes Discontinuous Shear Thickening is not directly affected by the local shear rate or gravity-induced inhomogeneities.  In Sec.~\ref{sec:normalstress} we show shear and normal stress measurements with different boundary conditions.  We show that the shear stress is coupled to the normal stress as in a frictional constitutive relation, rather than viscous, and that the rheology is controlled by the normal stress boundary condition.  Combining this with the shear profile measurements, we come to the surprising conclusion that for concentrated suspensions such as cornstarch in water which exhibit the phenomenon of Discontinuous Shear Thickening, the local constitutive relation between stress and shear rate is not actually shear thickening.   In Sec.~\ref{sec:dilation} we show images and movies of a visible change of the suspension-air boundary as particles penetrate the liquid-air interface in response to dilation of the granular packing under shear, which coincides with the shear thickening regime.  In Sec.~\ref{sec:surfacetension} we propose that this change in boundary condition due to dilation results in a confining stress from surface tension at the liquid-air interface, providing the normal stress boundary condition required for stress jump that characterizes Discontinuous Shear Thickening.  We then show that comparisons of the measured stress and dilation are consistent with this model, and that the confining stress scale from surface tension agrees with measurements of the maximum stress in the shear thickening regime $\tau_{max}$ for a wide variety of suspensions.   In Sec.~\ref{sec:wall} we generalize this result to other sources of confining stress  with measurements from a rheometer with solid walls instead of a liquid-air interface at the boundary and show that $\tau_{max}$ is generally set by the stiffness of the boundary.   Finally, in Sec.~\ref{sec:discussion} we discuss a generalization of the stress scales that bound the shear thickening regime to parameter regimes where other forces are relevant, the general constitutive relation, connections to the physics of other types of materials, and summarize the conditions for the occurrence of Discontinuous Shear Thickening.


\section{Background}
\label{sec:background}

Because of the vast amounts of literature referring to different phenomena and mechanisms which all fall under the category of shear thickening, we first carefully define what we mean by Discontinuous Shear Thickening.  We will suggest based on a literature review that hydrodynamic models which have been successful for describing other types of shear thickening have been insufficient for describing Discontinuous Shear Thickening.  Rather, there is significant evidence suggesting that dilation and normal forces play an important role which leads us to look for sources of stress similar to those in granular systems. 


Discontinuous Shear Thickening \citep{MW58, Ho72, Ho74, Ho82, Ba89, BLS90, Laun94, FHBM96, BW96, OM00, MW01a, MW01b, MW02, BBS02, LDH03, LDHH05, EW05, SW05, ENW06, LW06, FHBOB08, BJ09, BFOZMBDJ10, BZFMBDJ10} can be characterized by a set of several properties, which are observed to be similar in both dense suspensions and colloids.  Thus we find it convenient to define the phenomenon of Discontinuous Shear Thickening based on these properties and to distinguish it from other types of shear thickening:

\begin{enumerate}
\item \underline{Stress scales}:  The boundaries of the shear thickening regime are simply described in terms of stress scales (rather than shear rate) which are mostly independent of packing fraction  \citep{MW01a, SW05, BJ09} and liquid viscosity \citep{BLS90, FHBM96}.  Consequently, the onset shear rate varies with packing fraction and liquid viscosity since suspension viscosities increase with both parameters \citep{BB85}.  Shear thinning or Newtonian behavior is found at stresses below the onset of shear thickening at a stress $\tau_{min}$.  There is an upper bound of the shear thickening  regime at a stress $\tau_{max}$, and shear thinning is usually found at higher stresses.  

\item \underline{Diverging slope}: The term `Discontinuous' refers to the apparent discontinuous jump in the stress $\tau(\dot\gamma)$ of orders of magnitude in the shear thickening regime.  This jump is only observed at very high particle packing fractions $\phi$ (around 0.5 for nearly spherical particles).  The slope of $\tau(\dot\gamma)$ is only very steep over a small range in packing fraction (a few percent), lessening significantly at lower packing fractions \citep{MW01a, EW05, BJ09}.  The packing fraction dependence of the slope of $\tau(\dot\gamma)$ can be characterized as a power law diverging at a packing fraction $\phi_c$ \citep{BJ09}.  For $\phi > \phi_c$ the system is jammed meaning it will not flow for applied stresses below a non-zero yield stress of scale $\tau_j$.

\item \underline{Reversibility}:  The Discontinuous Shear Thickening described above is reversible, meaning viscosity curves are similar whether they are measured with increasing or decreasing stress histories.  Some examples of dramatic shear thickening have been found to be irreversible because of chemical-attraction-induced aggregation \citep{OKW08, LKZW10} or occur only in transient behavior \citep{FLBBO10}, and will not be considered here as they may be different phenomena.

\end{enumerate}

Because the above features of Discontinuous Shear Thickening are similar in both the suspension and colloid regimes, we refer to data in both regimes even though the dominant forces may be different.  

To facilitate an understanding of Discontinuous Shear Thickening, it may be useful up front to appreciate a disconnect between local and global viewpoints of rheology.  Standard rheology experiments measure a global mechanical response based on the drag force required to move two solid surfaces at some speed relative to each other with a fluid in between.  This is in contrast to the hydrodynamic theory of rheology which is based on continuum equations consisting of local constitutive relations between shear stress and shear rate.  One of the surprising results presented in this paper is that the global and local viewpoints of rheology lead to drastically different stress vs.~shear rate relations in the Discontinuously Shear Thickening suspensions being investigated.  Historically, Discontinuous Shear Thickening has been reported in experiments based on measurements of the global mechanical response.  We will show in Sections~\ref{sec:shearprofile} that the constitutive relation based only on the local shear rate can actually be {\em shear thinning} instead of shear thickening for these suspensions.  To resolve this apparent contradiction, in Sec.~\ref{sec:normalstress} we will show that most of the shear stress is due to frictional interactions which depend on the normal stress and can be controlled by the boundary conditions rather than the local shear rate.  Because the two standard viewpoints disagree on whether these dense suspensions are shear thickening or shear thinning, there will be some difficulty with terminology.  We chose to continue using the terminology of Discontinuous Shear Thickening to keep the connection to the previous literature, where the term is used extensively for similar experimental results.  We have chosen to capitalize `Discontinuous Shear Thickening' to refer to it as a name for a phenomenon rather than a local description, since from the local viewpoint it would not qualify as shear thickening.  Towards our goal of understanding Discontinuous Shear Thickening, our approach is to investigate all sources of the measured stresses, rather than just the hydrodynamic contribution to local constitutive relations.  


In contrast to the Discontinuous Shear Thickening described above, there are other types of shear thickening with different characteristics that have been modeled in hydrodynamic terms.  In the hydrocluster model, shear thickening occurs when an increasing shear rate leads to particle clustering and consequently increasing viscous stresses \citep{BB85, WB09}.  The shear thickening found in this model is relatively very weak in the sense that the viscosity increases by only a few percent per decade of shear rate \citep{BB85, MB04a}.  Inertial effects can also result in shear thickening, described in terms of either a Reynolds number or Bagnold number, in which inertial stresses can be characterized by $\tau(\dot\gamma) \propto \dot\gamma^2$ in the limit of high shear rate \citep{Ba54}.  In each of these cases, the onset of shear thickening is characterized by a shear rate, in contrast to an onset stress for Discontinuous Shear Thickening.   Because of the less steep $\tau(\dot\gamma)$, such shear thickening is often called Continuous Shear Thickening.  Continuous Shear Thickening can be observed even at very low packing fractions and has a much weaker packing fraction dependence than Discontinuous Shear Thickening.  In Sec.~\ref{sec:inertial} we will show an example of inertial shear thickening and characterize the parameter regime where it occurs in so we can clearly separate Discontinuous Shear Thickening from inertial stresses.  



Because of the qualitative differences between Discontinuous and Continuous Shear Thickening, we suspect they are different phenomena with different  mechanisms, and focus here only on Discontinuous Shear Thickening \footnote{In some of the literature, suspensions and colloids which exhibit discontinuous shear thickening at high packing fractions are said to exhibit continuous shear thickening at lower packing fractions because the slope of stress vs.~shear rate is lower.   However, there is usually no qualitative change in behavior or identifiable transition in scaling when the packing fraction is varied.   Thus we do not follow this convention because it does not suggest different phenomena and refer to such systems as exhibiting Discontinuous Shear Thickening.}. For this reason, we will not assume that results that apply to Continuous Shear Thickening also apply to Discontinuous Shear Thickening.  This is a different approach from much of the literature which has attempted to apply the hydrocluster model to both Continuous and Discontinuous Shear Thickening \citep{BB85, OM00, MW01a, GZ04, OKW08, WB09}.  The idea with the earlier approach has been to start with the hydrodynamic models of Continuous Shear Thickening which are well-understood at low packing fractions \citep{BBV02}, and extend them to higher packing fractions.  The expectation is that at higher packing fractions, hydrodynamically-induced clusters of particles form in which nearby particles act transiently as a solid cluster when they get so close to each other that the lubrication drag force between them blows up \citep{BB85, FMB97}.  While this mechanism seems plausible, so far the calculations have failed to reproduce steep viscosity curves comparable to experimental measurements of Discontinuous Shear Thickening.  

The major success of the hydrocluster model for Discontinuous Shear Thickening is the calculation of the stress at the onset of shear thickening $\tau_{min}$.  In early models for Brownian-motion dominated colloids the onset was described by a critical Peclet number $Pe = 6\pi\eta\dot\gamma a^3/kT$ for a particle size $a$ and thermal energy $kT$.  Shear thickening was expected to occur for $Pe \gg 1$ as the shear stress overcomes thermal diffusion of the particles \citep{BB85, MB04a}.  However, observations found the onset of shear thickening to be determined by the same stress at different packing fractions rather than the same shear rate.  This model, when using the suspension viscosity to convert to a stress scale $\tau_{min} = \eta\dot\gamma = kT/6\pi a^3$, has been successful at calculating the onset of both Continuous and Discontinuous Shear Thickening in Brownian colloids \citep{GZ04, MW01b}.  For colloids where electrostatic repulsions from a zeta potential $\zeta$ are dominant the above model had to be modified \citep{MW01a}.  In that case, the particular scaling found corresponds to a stress $\tau_{min} \sim \epsilon \zeta^ 2/a^2$ for permittivity $\epsilon$ which characterizes the electrostatic interactions between neighboring particles.\footnote{While the forces were calculated using a hydrodynamic model with particles interacting by a lubrication force, the hydrodynamic terms cancel up to a dimensionless coefficient of order 1.  Since the model was an order-of-magnitude calculation, it would have resulted in just as good a fit with the data if the interactions forces were calculated for a different particle separation distance or with a different mediating mechanism.}  In each regime, the modifications to the hydrodynamic model required to fit it to the data resulted in completely eliminating any dependence on hydrodynamic parameters such as viscosity or shear rate.  The resulting onset stress scale is not restricted to hydrodynamic mechanisms, as any type of force transferred through a continuum system can be expressed in terms of a stress.  Not surprisingly, with several relevant forces in colloids and suspensions, each of which could be dominant in different cases, a variety of different scalings for the onset stress have been found.  Depending on the parameter range, this dominant force could be Brownian motion \citep{BBV02, GZ04}, zeta potential \citep{MW01a}, induced dipole attractions \citep{BFOZMBDJ10}, or steric repulsion \citep{Ho98}.  Notably, in each case hydrodynamic terms such as shear rate and viscosity were found to be absent from  the modified scalings required to match the experiments.  This suggests inertia or hydrodynamics-based models are not necessary to determine the onset of Discontinuous Shear Thickening as initially envisioned by the hydrocluster model.  

Another observation that the hydrodynamic models require modification to describe is the upper stress boundary of the shear thickening regime $\tau_{max}$, since for inertial effects the viscosity increases monotonically with shear rate.  It has been suggested that this could be fixed by accounting for the finite stiffness of particles \citep{KMMW09}, but again this introduces a stress scale that is not necessarily a hydrodynamic in origin.  We will address the issue of the upper stress bound $\tau_{max}$ in Sec.~\ref{sec:surfacetension}.  

We could simply approach the problem in terms of stress scales as a modification of hydrocluster models as many others have done \citep{MW01a, OM00, GZ04, OKW08}.  Instead, here we focus on the scaling laws for the stress scales $\tau_{min}$ and $\tau_{max}$ to gain insight into the relevant physical mechanisms and come up with a description that encompasses all of these scalings without the need to refer to a hydrodynamic model.  A major advantage of this approach is that different mechanisms such as interparticle forces, gravity, and surface tension can be simply expressed in terms of stress scales which can be compared to the measured stresses without the need to reference a base model.    Another good reason to try to understand Discontinuous Shear Thickening without reference to hydrodynamic models is that some features of the rheology suggest a granular rather than hydrodynamic mechanism.  While a granular point of view is not necessarily the only way to understand the phenomenon, it will allow us to easily interpret many features.

Granular materials can have properties of solids, liquids, or gases under different conditions \citep{JNB96}.  For example, randomly packed particles at high enough packing fractions cannot shear or compress without deforming particles because geometric constraints force them to be in contact, so they have a yield stress like a solid.  At lower packing fraction the particles are able to move around each other freely in a liquid-like state.  The transition between these two regimes is sharp and is known as the jamming transition \citep{LN98, OSLN03}.  For shear thickening suspensions, the divergence of the slope of $\tau(\dot\gamma)$ at a critical packing fraction $\phi_c$ was found to correspond to the jamming transition \citep{BJ09}.  

Forces tend to be transmitted through jammed granular packings along concentrated paths called force chains \citep{CWBC98} such that the distribution of forces is characterized by an exponential tail \citep{MJN98, MB05, CJN05}.  Simulations of shear thickening colloids have similarly found contact networks between particles with an exponential distribution of forces under shear \citep{MB04b}.  



A feature of granular shear flows with special relevance to shear thickening is dilation \citep{Re1885, OL90}.  When a granular packing is sheared, the particles have to move around each other so the packing dilates, taking up more volume than it does at rest.  It has long been known that dilation occurs along with Discontinuos Shear Thickening (see \citet{MW58}, and references therein).    In fact, in some of the literature `dilatancy' has been used as a synonym for shear thickening \citep{Ba89}.  Especially important in understanding this relationship was the paper by \citeauthor{MW58}.  They showed that for suspensions of 0.2-1 $\mu$m TiO$_2$ particles, dilation initiated at stresses close to the onset of shear thickening for a range of packing fractions.  However, they found suspensions of 28-100 $\mu$m glass spheres in sucrose solutions dilated but did not shear thicken, showing that dilation was not always equivalent to shear thickening.  For 40+ years following this result, many of the major papers on shear thickening dropped the focus on dilation as a mechanism in favor of viscous mechanisms \citep{Ho82, BB85, MW01a}.   

However, there is another possible interpretation of the data presented by \citet{MW58}.  They confirmed for several suspensions that the onset of shear thickening coincided with dilation, as had been seen in many previous results (see references in \citet{MW58}).  Taken together with the observation of dilation in the absence of shear thickening, this inductively suggests that dilation is necessary but not sufficient for Discontinuous Shear Thickening.  More recent results have shown that shear thickening can be hidden by a yield stress or other shear thinning effect \citep{BFOZMBDJ10}.  Specifically, \citeauthor{MW58} used glass beads ranging from 28-100 $\mu$m in diameter in a Couette geometry, and such large, heavy particles will jam in a Couette cell because they settle under gravity, resulting in a yield stress \citep{FBOB09}.  This can explain why shear thickening was not observed for the settling particles used by \citeauthor{MW58}.  It has only been in the last 10 years that dilation has become prominent again in the shear thickening literature \citep{OM00, LDH03, LDHH05, FHBOB08}.   However, a mechanism by which dilation leads to the dramatic increase in stress in Discontinuous Shear Thickening has yet to be explained.  This is the subject of Sections \ref{sec:dilation}, \ref{sec:surfacetension}, and \ref{sec:wall}.

When dilation of granular shear flows is prevented by confinement, shear is instead accompanied by normal forces against the walls \citep{Re1885, OL90}.  Dilation of packings against boundaries can play a dominant role in the mechanics of granular systems where confining pressures from the boundary are transmitted through the material \citep{LW69}.  In Discontinuous Shear Thickening suspensions normal forces are usually found to be positive, meaning the sample is pushing against the rheometer plate as expected for dilation \citep{JR93, LDH03, LDHH05, FHBOB08, BZFMBDJ10}.  It was proposed by \citet{FHBOB08} that shear thickening cannot occur if the normal stress is taken away.  We will test this in Sec.~\ref{sec:normalstress}. 

\section{Materials and methods}
\label{sec:methods}

\subsection{Suspensions}

We studied a wide variety of suspensions with different particle sizes and shapes, liquid viscosities, and density differences among other properties to investigate the mechanism for Discontinuous Shear Thickening.  We use this variety as a way to determine which features are common to all of the suspensions that exhibit Discontinuous Shear Thickening. 

As a prototypical shear thickener we used cornstarch obtained from Argo.  Cornstarch particles have a mean diameter of 14 $\mu$m and density of 1.59 g/mL based on buoyancy in CsCl solutions.  They are very hydrophilic and hard -- with a compression modulus on the order of $10^{10}$ Pa -- at room temperature.  At higher temperatures the polymers that compose cornstarch particles can gel.  To compare suspensions with different liquid viscosities, we suspended cornstarch in either a mixture of 61.5\% water and 38.5\% CsCl by weight with a viscosity of 1 mPa$\cdot$s and density of 1.41 g/mL or a mixture of 73.5\% glycerol 13.0\% water, and 13.5\% CsCl by weight with a viscosity of 80 mPa$\cdot$s and density of 1.34 g/mL.  For such small particles, the settling time is several hours even without density matching.  

For a series of suspensions in which we varied particle size we used soda-lime glass spheres with a density of 2.46 g/mL.  We obtained particles with nominal diameter ranges of 3-10 $\mu$m, 10-25 $\mu$m, and 15-40 $\mu$m from Corpuscular, 45-63 $\mu$m, 75-104 $\mu$m (referred to as 100 $\mu$m), 177-250 $\mu$m, and 400-595 $\mu$m from MoSci (Class IV), and 1120-1350 $\mu$m and 1900-2100 $\mu$m.  For polydisperse suspensions, measurements of $\tau_{min}$ could be collapsed onto the same curve as monodisperse suspensions \citep{MW01a}, suggesting that polydispersity does not need to be accounted for, so we will compare particle distributions only by their mean diameter.   The glass spheres were dispersed in various liquids, including water or mineral oil with a viscosity of 58 mPa$\cdot$s and density of 0.87 g/mL.   

For visualization purposes we used two types of opaque particles.  The first, referred to as ZrO$_2$, were spheres obtained from Glen Mills consisting of 69\% ZrO$_2$ and 31\% SiO$_2$.  They have a nominal diameter range of 100-200 $\mu$m and a density of 3.8g/mL. These particles were dispersed in the same mineral oil used for the glass particles. For experiments with density matched suspensions, we used polyethylene spheres obtained from Cospheric.  These particles have a nominal diameter range of 125-150 $\mu$m and density of 1.01 g/mL.  They were dispersed in silicone oil AR 20 with a nominal density of 1.01 g/mL and viscosity of 20 mPa$\cdot$s.  When varying the temperature of the suspension, we found that the settling time was minimized at 19$^{\circ}$ C from which we estimated a density difference of order $10^{-4}$ g/mL.

We also include some summary data for particles fabricated into different shapes from polyethylene glycol (PEG) suspended in liquid PEG-250 from \citet{BZFMBDJ11}, and 100 $\mu$m spheres made of polystyrene dimethyl ether suspended in PEG.

\subsection{Tools}

\begin{figure}                                                
\centerline{\includegraphics[width=3.8in]{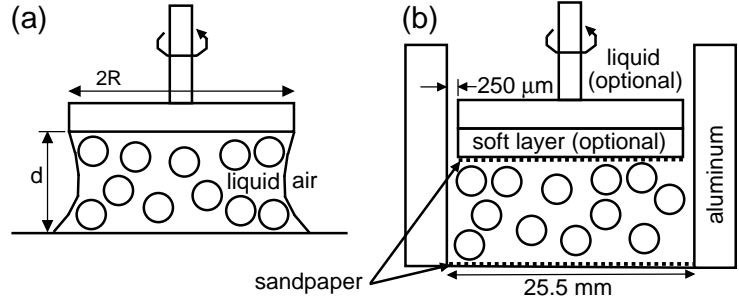}}
\caption{(a) A standard parallel plate rheometer setup.  The suspension is confined between the plates by the surface tension of the suspending liquid.  (b) A modified parallel plate setup with solid walls around the sample.  In this setup the use of a suspending liquid is optional, and the wall stiffness can be modified by inserting layers of different stiffness between the suspension and plate.}
\label{fig:rheometer}                                        
\end{figure}

Measurements were performed with an Anton Paar Physica MCR 301 rheometer which measures the torque $T$ required to shear a sample at a tool angular rotation rate $\omega$.    Most measurements were done in a parallel plate setup where normal forces could be measured.  This geometry is shown in Fig.~\ref{fig:rheometer}a and characterized by the plate radius $R=12.5$ or $25$ mm and gap size $d$ between the plates.  A few measurements were done with a cylindrical cup-and-bob (Couette) geometry in which the environment is better controlled.  The tool surfaces are smooth stainless steel.  The viscosity, indicating the mechanical resistance to shear, is defined as $\eta\equiv\tau/\dot\gamma$ in a steady state.  For the parallel plate setup, for example, we represent the global shear stress by

\be
\tau = \frac{2T}{\pi R^3}
\label{eqn:stress}
\ee

\noindent and shear rate by

\be
\dot\gamma = \frac{R\omega}{d} \ .
\label{eqn:shearrate}
\ee

\noindent  These standard coefficients in equations \ref{eqn:stress} and \ref{eqn:shearrate} correspond to the values at the outer radius of the plate in the case of a Newtonian shear profile.  Since the dramatic feature of Discontinuous Shear Thickening is the increase in stresses under shear, the reported global shear stress and shear rate values are meant to characterize this response, in which the viscosity is a measure of mechanical energy dissipation, rather than a local constitutive relation between shear stress and shear rate.  We will show in Sec.~\ref{sec:shearprofile} that the local and global constitutive relations are quite different for Discontinuous Shear Thickening suspensions.  The globally averaged stress is still appropriate to characterize forces in a way that is independent of system size so that measurements can be compared to the scale of forces associated with different physical mechanisms.  Since most of the suspensions used are not density matched and non-Brownian, they can be very inhomogeneous.   In this case, the given stress and shear rate relations may overestimate the average values by as much as 25\% depending on the inhomogenieties.  Since the formalism of local hydrodynamic constitutive relations and normal stress differences assumes homogeniety, we will not assume it applies, and instead base our interpretations on the gross mechanical response.



In the parallel plate measurements, the upward force on the rheometer tool is measured and the mean normal stress $\tau_N$ is obtained by dividing this normal force by the plate cross-sectional area.  The standard deviation of the force measured during a static measurement over 10 s intervals with or without sample is $6\times 10^{-4}$ N, giving an uncertainty of 0.3 Pa (1.2 Pa) for a 50 mm (25 mm) diameter plate within a single measurement run.  The standard deviation of the average force measured after calibration with no sample is $4\times10^{-3}$ N, giving an uncertainty of 2 Pa (8 Pa) for a 50 mm (25 mm) diameter plate when comparing different runs.  The resolution of the shear stress is much better, with an absolute uncertainty less than (0.001 Pa) 0.01 Pa for a 50 mm (25 mm) plate.

The Anton Paar MCR 301 rheometer has special settings for normal force control measurements. Our reported measurements were done with the value of `normal force hysteresis' set to 0.001 N.  This value controls how much the normal force can deviate from the set value before the plate moves in response, although in practice the plate tends not to move until the normal force deviation exceeds about 0.01 N (20 Pa for the 25 mm plate).  The `normal force dynamics' value was set to 0\% (default).   This value controls the acceleration of the gap size in the feedback loop.  Under this setting the gap responds slowly to variations in the normal force, and produces the most reproducible steady state gap sizes.

The typical parallel plate setup is shown in Fig.~\ref{fig:rheometer}a, in which the suspension is held in place between the parallel plates by surface tension at the liquid-air interface around the side.  Because we saw that particles could penetrate the liquid-air interface (see Sec.~\ref{sec:dilation}) which we suspected could modify the stresses on the suspension from the interfacial tension, we desired a different boundary condition with a hard wall for some experiments.  To accomplish this we machined an aluminum cylindrical cup with inner diameter 25.5 mm as shown in Fig.~\ref{fig:rheometer}b.  The cup fits around the tool with a gap small enough to prevent 500 $\mu$m diameter particles from slipping through but large enough to allow the tool to rotate without friction.  This cup confined the particles to the volume beneath the plate, while the liquid could be filled to a higher level or omitted altogether so there was no liquid-air interface for particles to penetrate.   The plates were covered with sandpaper sheets with a grit size of 100 $\mu$m to avoid slip with dry grains.  In some cases we inserted a soft layer in between the top plate and the sandpaper to modify the compliance of the wall.

\subsection{Measurement procedure}

Packing fractions $\phi$ were calculated as the volume of solid particles over the total volume of particles plus liquid mixed together.  The packing density in terms of the inverse of the available free volume per particle may decrease slightly during measurements as the grain packing dilates.  Above the jamming transition, this packing density may also be less than the measured packing fraction if air bubbles become trapped in the interior.  Humidity also has a large effect on the amount of water adsorbed onto dry grains open to the atmosphere, especially cornstarch which is so hygroscopic that 10-20\% of the weight of the `dry' powder is from water.  Comparison of density measurement techniques suggest that cornstarch is porous or that it may absorb CsCl, thus we report mass fractions $\phi_m$ for cornstarch rather than volumetric packing fractions.  



Measurements were made with the rheometer's bottom plate temperature controlled at $20^{\circ}$ C.  The room humidity ranged from 22\% to 38\%, although during individual experiments the humidity was constant.  To minimize evaporation or adsorption of water from the atmosphere to the suspension during measurements, we used a solvent trap when the suspending liquid was water which enclosed the sample and a small amount of air around it by an extra layer of liquid.  The enclosed air equilibrated with the sample to prevent further changes to the suspension.  

The gap size $d$ for parallel plate measurements was usually about 1 mm, large enough to avoid finite-size effects on the viscosity for particles around 100 $\mu$m in diameter \citep{BZFMBDJ10}.  We measured bulk shear thickening with both rough and smooth plates and did not find any difference in the shear thickening due to the plate surface.  To directly measure slip, we used video microscopy to observe the shear profile at the outer edge of the plate.  The results of these measurements are shown in Sec.~\ref{sec:shearprofile}.  We visually confirmed for all reported measurements that the suspensions do not spill.  Spillage was often the limiting factor in the maximum stress or shear rate applied for our measurements.

\begin{figure}                                                
\centerline{\includegraphics[width=5.5in]{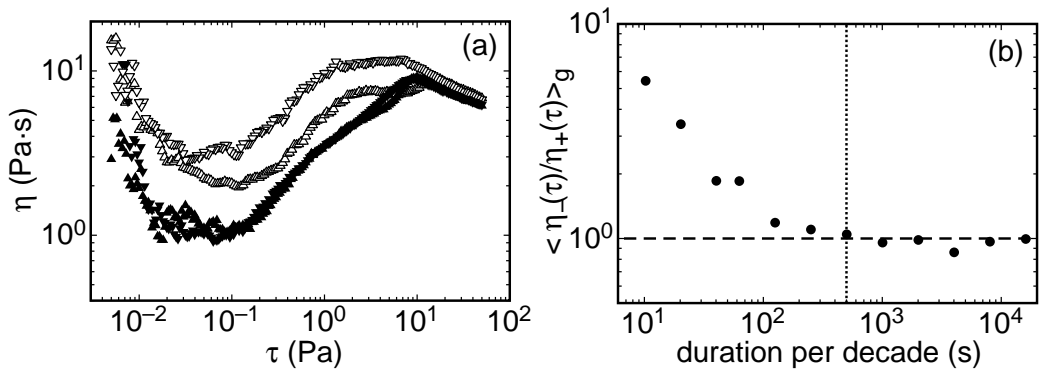}}
\caption{(a)Apparent viscosity curves for 100 $\mu$m glass spheres in mineral oil taken with different measurement durations to show hysteresis loops.   Open symbols: measurement duration of 40 s per decade of the stress ramp.  Solid symbols: 500 s per decade.  Up-pointing triangles correspond to increasing stress ramps, while down-pointing triangles correspond to decreasing stress ramps.  (b) Characterization of the hysteresis as the geometric mean of the viscosity ratio between the decreasing and increasing ramps of the hysteresis loop, plotted for different ramp durations per decade of stress.  Dashed line:  a ratio of 1 between increasing and decreasing ramps corresponding to no hysteresis. Dotted line: ramp rate used for later steady state measurements 
}
\label{fig:hysteresis}                                        
\end{figure}

Suspensions were first pre-sheared immediately before measurements for at least 100 seconds at shear rates above the shear thickening regime where the steady state flow is fully mobilized, then viscosity curves were measured by ramping the control parameter (shear stress or rate) down and then up to obtain hysteresis loops.  To ensure that we obtain steady state viscosity curves, the measurement ramp should be long enough that the size of the hysteresis loop is equal to that of the infinite duration limit.  To check this, we show data in Fig.~\ref{fig:hysteresis} for a sample of 100 $\mu$m glass spheres in mineral oil at $\phi=0.56$, which is a stable sample over long time periods because the oil does not evaporate.  Viscosity curves shown were taken first with a decreasing stress ramp followed by an increasing stress ramp for several different ramp durations.  Since the control is a logarithmic ramp in stress over 4 decades, the measurement duration is specified in terms of duration per decade of stress.  The hysteresis effect is characterized by the average distance between the upper and lower branches of the hysteresis loop in $\eta(\tau)$ on a log-log scale.  This is calculated equivalently as the geometric mean of viscosity ratio $\langle \eta_-(\tau)/\eta_+(\tau)]\rangle_g$ where $\eta_-$ and $\eta_+$ are viscosities for decreasing and increasing stress ramps, respectively, and $\langle...\rangle_g$ indicates a geometric mean, i.e. averaged on a log-log scale.  This average was done over the stress range of 0.1 to 8 Pa in the shear thickening regime.  The hysteresis initially decreases with increasing measurement duration, then levels off for long measurements indicating a steady state limit.  The initial duration-dependent behavior is characteristic of a transient relaxation, and the crossover between the regimes indicates a characteristic timescale for the sample to reach steady state.   The leveling off of the viscosity ratio at a value of 1 suggests that in this case there is not a true hysteresis effect.  At packing fractions very close to the jamming transition we sometimes find some non-zero hysteresis loop even for very long measurements, such that different steady states can be reached dependent on the shear history.  The steady state viscosity curves we report are generally in the long-duration regime where the viscosity ratio has leveled off, for this sample we use a control ramp rate of 500 s per decade of stress.   For each steady state measurement, we ramped the control parameter down then up at least once, and in some cases up to five times, but we show only one set of curves for brevity if they were all identical within typical variations of 10-20\% from run to run. 


\section{Inertial scalings}
\label{sec:inertial}

In this section we show some examples of viscosity curves for suspensions at different packing fractions.  While there are already many examples of the packing fraction dependence, here we vary the liquid viscosity and the packing fraction from near the jamming transition all the way down to zero to compare to hydrodynamic scalings that apply to suspensions at low concentrations.  

\begin{figure}                                                
\centerline{\includegraphics[width=6.5in]{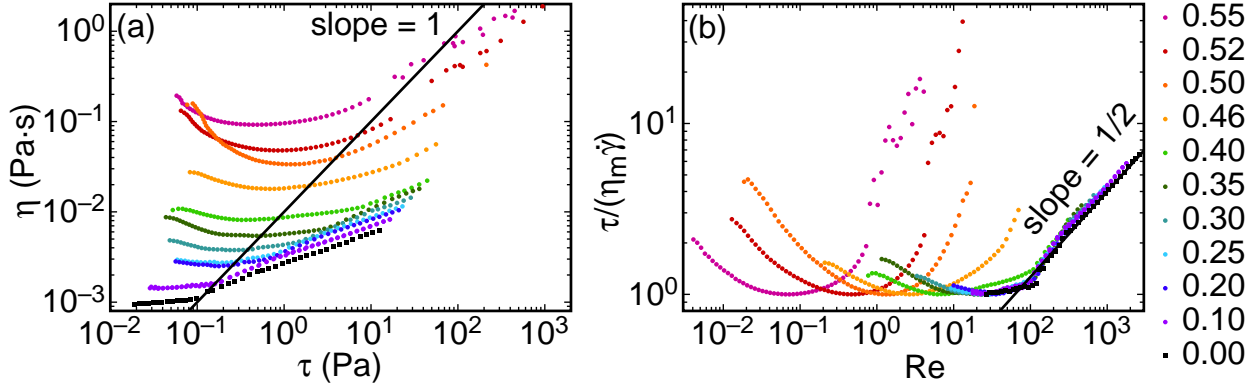}}
\caption{(a)  Viscosity vs. stress curves for suspensions of cornstarch in water.  Mass fractions $\phi_m$ are shown in the key; higher curves correspond to larger $\phi_m$, and $\phi_c=0.56$.  The solid line corresponds to a constant shear rate.  (b) A rescaling of the data as $\tau/\eta_m\dot\gamma$ vs. Reynolds number $Re = \rho_l\dot\gamma d^2/\eta_m$ which should result in a data collapse for hydrodynamic flows.   The line corresponds to a scaling $\tau\sim \dot\gamma^{3/2}$ in the range $100  \stackrel{<}{_\sim} Re  \stackrel{<}{_\sim} 3000$.}
\label{fig:reynolds}                                        
\end{figure}

We first show measurements of cornstarch suspended in water.  These measurements were made with the Couette geometry.  Viscosity is plotted vs.~shear stress for several packing fractions in Fig.~\ref{fig:reynolds}a.  Apparent shear thickening is seen as regions with a positive slope of the viscosity curve for all packing fractions, including $\phi_m=0$ which is pure water.  For pure liquids this behavior can be quantified in terms of a dimensionless Reynolds number, which represents a ratio of inertial to viscous stresses.  This Reynolds number is usually of the form $\rho_l d^2\dot\gamma/\eta_l$ for pure liquids where $\rho_l$ and $\eta_l$ are the density and dynamic viscosity of the liquid, respectively.  Thus, for the pure liquid the transition from a viscous-dominated regime with a nearly constant viscosity to an inertia-dominated regime with apparent shear thickening occurs at a fixed shear rate.  However, we find that the sharp transition does not occur at the same shear rate for different packing fractions; specifically for low packing fractions the onset shear rate increases with packing fraction.  The contribution of viscous stresses to the viscosity of dense suspensions is much higher than that of the pure liquid \citep{BB85}, so the viscous term in the denominator of the Reynolds number should be modified for suspensions.  For the contribution of viscous stresses to suspension viscosity we do not take the zero shear rate limit of the viscosity since in this limit suspensions rheology can be dominated by non-viscous particle interactions which result in shear thinning at low shear rates \citep{Ba89, MW01a, BFOZMBDJ10}.   Rather, we take as our best estimate the minimum suspension viscosity $\eta_m$ which occurs at the onset of shear thickening for each packing fraction.  Our suspension Reynolds number is then $Re = \rho_l \dot\gamma d^2/\eta_m$.

In Fig.~\ref{fig:reynolds}b we plot the shear stress normalized by $\eta_m\dot\gamma$ (corresponding to the viscous contribution to the stress) vs. suspension Reynolds number $Re$.  With this non-dimensionalization, data for any Newtonian fluid should collapse onto the same curve, which is nearly flat at low $Re$ in the viscous regime and transitions to a more positive slope at higher $Re$ due to inertial effects.   We find that the lower mass fractions $\phi_m\stackrel{<}{_\sim} 0.4$ indeed  collapse onto a single curve.   The very mild increase in effective viscosity for $1\stackrel{<}{_\sim} Re \stackrel{<}{_\sim}100$ is typical of hydrodynamic flows in this range of $Re$ where viscous stresses are dominant but inertial effects start to become measurable \citep{Schlichting, KM08}.  The approximate scaling $\tau\sim \dot\gamma^{3/2}$ for $100 \stackrel{<}{_\sim} Re  \stackrel{<}{_\sim} 3000$ as inertia becomes stronger is also typical of hydrodynamic flows \citep{DS60}.  The asymptotic scaling $\tau\sim \rho\dot\gamma^2 d^2$ expected in the fully inertial regime is usually not found until $Re \stackrel{>}{_\sim} 10^3$ \citep{Schlichting}.  The data collapse and scaling for $\phi_m\stackrel{<}{_\sim} 0.4$ suggests that in this regime the suspension behaves like a Newtonian fluid.  In contrast, for $\phi_m \stackrel{>}{_\sim} 0.4$ the normalized viscosity curves deviate significantly from this collapse and scaling, increasing more steeply than inertial or viscous stresses are expected to be able to grow, and the steep shear thickening onsets at much lower values of $Re$.  The onset of shear thickening occurs at {\em lower} shear rates for larger $\phi$, instead of the {\em higher} shear rates which would be required before inertial stresses dominate over viscous stresses.\footnote{For $\phi_m \stackrel{>}{_\sim} 0.4$, $\eta_m$ is likely an overestimate of the viscous contribution to viscosity as the non-Newtonian terms become larger near the jamming transition and $\eta_m$ likely represents a cross-over between the shear thinning and shear thickening effects \citep{BFOZMBDJ10}  Regardless, since the viscous contribution to the viscosity increases while the onset shear rate decreases with packing fraction, the onset Reynolds number still becomes very low at high packing fractions.}  Instead, the onset appears to be set by a constant stress scale for $\phi_m \stackrel{>}{_\sim} 0.4$ as seen in Fig.~3a.  This suggests that inertial stresses are relatively small in this regime where shear thickening becomes more dramatic. This does not necessarily imply that viscous forces are dominant in the system, rather the lack of data collapse by the hydrodynamic non-dimensionalization for $\phi_m \stackrel{>}{_\sim} 0.4$ in Fig.~\ref{fig:reynolds} suggests other stresses must be involved.   


\begin{figure}                                                
\centerline{\includegraphics[width=3.in]{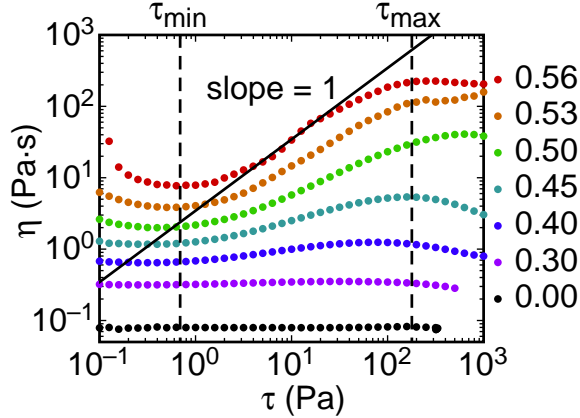}}
\caption{Viscosity curves for cornstarch in a glycerol-water mixture in which the Reynolds number remains small.  Mass fractions $\phi_m$ are shown in the key; higher curves correspond to larger $\phi_m$, and $\phi_c=0.58$.  The solid line of slope 1 corresponds to a constant shear rate and the steepest possible steady state viscosity curve.  The vertical dashed lines define the stress scales $\tau_{min}$ and $\tau_{max}$ that bound the shear thickening regime.}
\label{fig:glycerol}                                        
\end{figure}

The scaling seen in Fig.~\ref{fig:reynolds} suggests two competing mechanisms for different types of shear thickening -- inertial and Discontinuous -- such that only the stronger effect is observed in a single viscosity curve.  To obtain a system where the Reynolds number remains low in the Discontinuous Shear Thickening stress range even at low packing fractions, a liquid of higher viscosity can be used.  Accordingly, we suspended cornstarch in a glycerol-water mixture with a viscosity 80 times that of water.  The viscosity curves vs. stress for different packing fractions can be seen in Fig.~\ref{fig:glycerol}.  Comparing with the data for cornstarch in water in Fig.~\ref{fig:reynolds}a, we can see that there is similar strong shear thickening at high packing fractions, but no apparent shear thickening at low packing fractions.  At the same stress, the Reynolds number is lower by about a factor of the ratio of the viscosities squared ($\approx 600$) for the suspension with glycerol, thus $Re<100$ and inertial effects remain negligible in the stress range of Discontinuous Shear Thickening even in the limit of zero packing fraction.  We can see the remaining shear thickening uncontaminated by inertial effects is now very weak at $\phi_m=0.40$, and is almost imperceptible at $\phi_m=0.30$.  This is a typical example of Discontinuous Shear Thickening, in which the region with positive slope of $\eta(\tau)$ occurs in a stress range that is nearly independent of packing fraction.  This slope increases with packing fraction, approaching $\eta\sim \tau$ (solid line in Fig.~\ref{fig:glycerol}) corresponding to a discontinuous stress/shear-rate relation.  The bounds of the shear thickening regime are characterized on the lower end by $\tau_{min}$ defined as the onset of a positive slope of $\eta(\tau)$, and on the upper end by $\tau_{max}$ defined as the transition from positive to negative slope.  These transitions are measured as the crossover between local power law fits on either side.  Because of fluctuations typically on the scale of 10-20\% in the viscosity we do not count any features smaller than that threshold as distinct transitions.  

There are several other dimensionless numbers that have been used to describe inertial effects in particulate flows.     In particular, often the system size $d$ is replaced with the particle size $a$, for example leading to a Bagnold number \citep{Ba54} or particle Reynolds number.  The distinction between the two types of scalings can be made with a pure liquid at zero packing fraction which has the system size scale of $d$ but no particle length scale.  Since the cornstarch-in-water data for $\phi_m\stackrel{<}{_\sim} 0.4$, including $\phi_m=0$, collapse based on a Reynolds number scaling in terms of $d$, the system size should be taken as the relevant length scale as is typical in pure fluids, and a  dimensionless inertial number based on particle size would not be able to collapse the data over as wide a range.  

We can qualitatively distinguish inertial flows from Discontinuous Shear Thickening because the packing fraction, shear rate, liquid viscosity, and gap size dependence of inertial flows all differ from that of Discontinuous Shear Thickening, and the steepest possible scaling for inertial flows is $\tau\sim\dot\gamma^2$ in the limit of large shear rates \citep{Ba54}.  Since the focus of this paper is on Discontinuous Shear Thickening,  all of the following data will be in the high packing fraction regime (equivalent to $\phi_m \stackrel{>}{_\sim} 0.4$) and for $Re < 100$ to avoid mixing inertial effects with Discontinuous Shear Thickening. 



\section{Gravity and the onset stress}
\label{sec:onsetstress}

Suspensions and colloids occupy a region of phase space where many physical forces may be relevant; these include Brownian motion, gravity, surface tension, and electrostatics.  One consequence of this is that there are different scaling laws for the onset of shear thickening in different parameter regimes where one of these forces is dominant \citep{Ho98, MW01a, MW01b, BBV02, SW05, BFOZMBDJ10}.   Here we address the case of a gravity-dominated regime for large particles. This regime is not yet as well-characterized as the other regimes, but an understanding of the effects of gravity on the onset stress will make it possible to further generalize the conditions for the onset of shear thickening with a mechanism that accommodates all of these scalings.


\begin{figure}                                                
\centerline{\includegraphics[width=3.in]{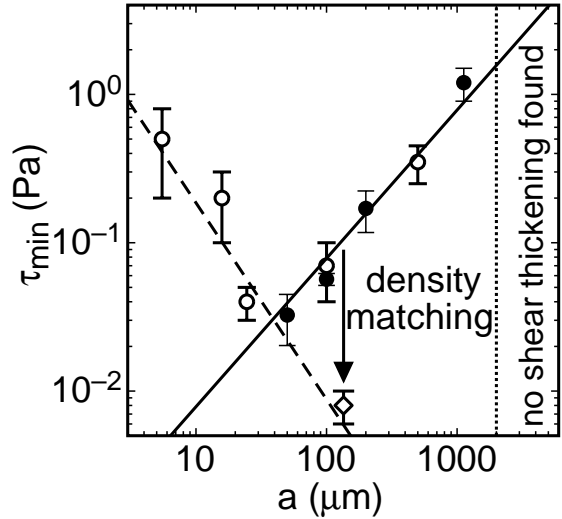}}
\caption{The stress at the onset of shear thickening $\tau_{min}$ for glass spheres of different diameters $a$ in mineral oil (solid circles, $\Delta\rho=1.58$ g/mL)  or water (open circles, $\Delta\rho=1.46$ g/mL).  Open diamond:  polyethylene in silicone oil ($\Delta\rho \approx 0.0001$ g/mL).  The solid line is the shear stress required to lift particles off the top layer of the packing against friction and gravity $\mu_{eff}\Delta \rho g a/15.3$.  Dashed line: representative curve for data where gravity is not the dominant interparticle interaction. Dotted line: bound above which larger particles did not exhibit any shear thickening regime.
}
\label{fig:sizestressgravity}                                        
\end{figure}

We measured steady state viscosity curves for a series of suspensions of glass spheres ($\rho=2.46$ g/mL)  with diameters ranging from 6 to 2000 $\mu$m in diameter in either mineral oil ($\rho=0.88$ g/mL) or water ($\rho=1.00$ g/mL) at packing fractions ranging from 0.50 to 0.58  These particles are large enough to settle over time since the glass is much denser than the liquids.  In each case, the major features were qualitatively similar to Fig.~\ref{fig:glycerol}, with increasingly steep slopes at higher packing fractions and shear thickening occurring in a relatively fixed stress range.  We obtained mean values of $\tau_{min}$, corresponding to the onset of a positive slope of $\eta(\tau)$,  for each suspension averaged over a range of packing fractions close to but below $\phi_c$.  The mean values of $\tau_{min}$ for each particle size are plotted in Fig.~\ref{fig:sizestressgravity}. 



For the largest particles with $a\ge 500$ $\mu$m, the suspensions would not remain confined between the rheometer plates with a vertical boundary because the particles are so heavy that they can no longer be confined by surface tension.  Since this confinement is set by the interplay between gravitational and surface tension forces, it is no surprise that loss of confinement occurs for particles on the order of the capillary length, or $\sim 1$ mm.  We could still make measurements with some sample extended outside the area between the plates to obtain the scales of $\tau_{min}$ and $\tau_{max}$.   For the largest glass beads with a diameter of 2000 $\mu$m, we found no shear thickening regime.  The significance of this maximum particle size for shear thickening will be addressed in Sec~\ref{sec:stressscales}.

It can be seen in Fig.~\ref{fig:sizestressgravity} that there are two distinct scaling regimes for $\tau_{min}$ which meet at a minimum near a particle size of 50 $\mu$m.  For smaller particles with $a\le 50$ $\mu$m that approach the colloidal regime, interparticle interactions from various sources including electrostatics and Brownian motion tend to become large relative to gravity and can affect the onset stress.  One effect which is relevant here is a high particle-liquid interfacial tension which results in an effective attraction between particles which can form force chains that span the system and jam it.  This in turn results in a yield stress and shear thinning even at low packing fractions which then can hide shear thickening \citep{BFOZMBDJ10}. Specifically for this measurement series, with glass beads 50 $\mu$m and larger, the particles will disperse well and shear thicken in either oil or water.  However, the 6 $\mu$m glass particles are effectively hydrophilic.  Consequently in oil they have a significant yield stress and shear thickening was not observed at all.  

To understand the scaling of $\tau_{min}$ for the larger particles with $a\ge 50$ $\mu$m, we now analyze the effects of gravity on non-Brownian suspensions.  In the limit of zero shear rate, gravity results in particles settling and resting on the bottom plate.  The measured stress would come only from shear of the thin fluid layer on top of the settled particles.  The drag force from the shear in the liquid layer above the particles can start to move the upper layer of particles if it exceeds the static frictional force between particles under gravity.   In a parallel plate geometry, the horizontal cross-section has a uniform area so, to balance forces, the shear stress $\tau$ must be on average independent of height.  As an estimate for the drag force on a particle in the top layer, we use the drag force on a sphere sitting on a flat surface is $2.55\pi\tau a^2$ \citep{GCB67}.  The frictional force on one of these particles is $\pi \mu_{eff} \Delta\rho g a^3/6$ for an effective static coefficient of friction $\mu_{eff}$, density difference $\Delta\rho$ between the particles and liquid, and acceleration of gravity $g$.  Since the particles are sitting on a pile of beads instead of a flat surface, the effective friction is enhanced by geometry because of the need for the spheres to rise over the particles in the layer below.  To measure $\mu_{eff}$, we glued 1 mm glass beads in a monolayer to a glass slide.  We then performed an inclined plane test with this system immersed in water, setting individual glass beads on top of the bead-covered slide and slowly tilting the slide until the loose beads started falling down.  From this we obtained $\mu_{eff}=0.8\pm 0.1$.  Balancing the drag and frictional forces gives the stress at the onset of shear between grains to be $\mu_{eff}\Delta\rho g a/15.3$.  This prediction is plotted in Fig.~\ref{fig:sizestressgravity}.  It is seen to match well the measured onset stress $\tau_{min}$ for particles between 50 and 1125 $\mu$m in diameter.  This confirms that the onset of shear thickening in the gravity-dominated regime is set by the stress required to initiate shear of the particles against gravity and friction.  

Since the onset scaling for large particles is set by gravity, this suggests $\tau_{min}$ can be lowered by density matching.  We tested this by measuring steady state viscosity curves for 100 $\mu$m polyethylene particles in silicone oil with a density difference of about $10^{-4}$ g/mL.   The mean value of $\tau_{min}$ is shown in Fig.~\ref{fig:sizestressgravity} by the open diamond.   While $\Delta \rho$ was reduced by a factor of $10^4$ compared to the non-density matched case, the onset stress was only reduced by an order of magnitude.   In this case the density-matched data fall onto a similar scaling as was found for the glass for $a \le 25$ $\mu$m.  In many cases for particles even as large as 100 $\mu$m, we found density matching can have no effect on the onset stress due to the significance of interparticle interactions.  For cornstarch in water, density matching by adding CsCl to the water did not reduce the onset stress.  For glass spheres in a heavy liquid $\rho=2.46$ g/mL (Cargille labs inorganic salt series) we found no measurable decrease in the onset stress compared to mineral oil or water, and found shear thinning below the onset stress as opposed to the Newtonian scaling found for glass suspensions whose onset is determined by the gravitational scaling \citep{BJ09, BFOZMBDJ10}.  This suggests that in each of these cases, the stress scale characterizing interparticle interactions which is dominant for smaller particles is very close to the onset stress if not the dominant factor.  These results show that while density matching can lower the onset stress in the gravity-dominated regime, it cannot do so beyond the limits set by any other stress scales due to particle interactions.  Thus we generally expect a larger effect of density matching for very large particles further into the gravity dominated regime.  

In this set of experiments with a parallel plate setup, gravity caused particles to settle and shear thickening required initiating shear so the onset stress $\tau_{min}$ scaled like a hydrostatic pressure due to the weight of the top layer of particles only.  In contrast, in a Couette cell with vertical walls a yield stress was found that scales with the same hydrostatic pressure in the suspension pushing on the walls \citep{FBOB09}.  We know that such a yield stress can hide shear thickening if it is larger than the stress from shear thickening mechanisms \citep{BFOZMBDJ10}.  This implies that the effect of the yield stress from gravity is shear-geometry dependent because of the directionality of gravity.  The fact that this yield stress can move the onset stress and hide shear thickening if the yield stress is larger \citep{BFOZMBDJ10} suggests it works in addition to stresses responsible for shear thickening, and there is no indication that the shear thickening mechanism itself is affected by gravity.  We will revisit the issue of the significance of the scalings for the onset stress in Sec.~\ref{sec:stressscales} after a mechanism for shear thickening is identified.


\section{Shear profile}
\label{sec:shearprofile}

In this section, we show shear profile measurements of both density matched and non-density-matched suspensions that exhibit Discontinuous Shear Thickening.  The inhomogeneity due to gravity creates a shear gradient that allows us to separate out the contributions from viscous forces, gravity, and other forces to the constitutive relation between stress and shear rate in the shear thickening regime.  

To measure the shear profile we used a video camera with a bellows and magnifying lens to obtain a pixel size as small as 10 $\mu$m.  The camera was placed next to the standard parallel plate rheometer setup and focused on the outer edge of the sample in the plane of the shear direction and shear gradient.  While there is some distortion from looking through the curved liquid-air interface, we can track individual particle motions to measure the shear profile at the edge of the sample.  Videos were taken for constant shear rate conditions after the steady state was reached.  Steady state shear profiles obtained by using particle image velocimetry to obtain local particle velocities and averaging the velocities at each height.  A small tilt of the camera caused a smoothing effect over about 4\% in the depth. 
 
\begin{figure}
\centerline{\includegraphics[width=6in]{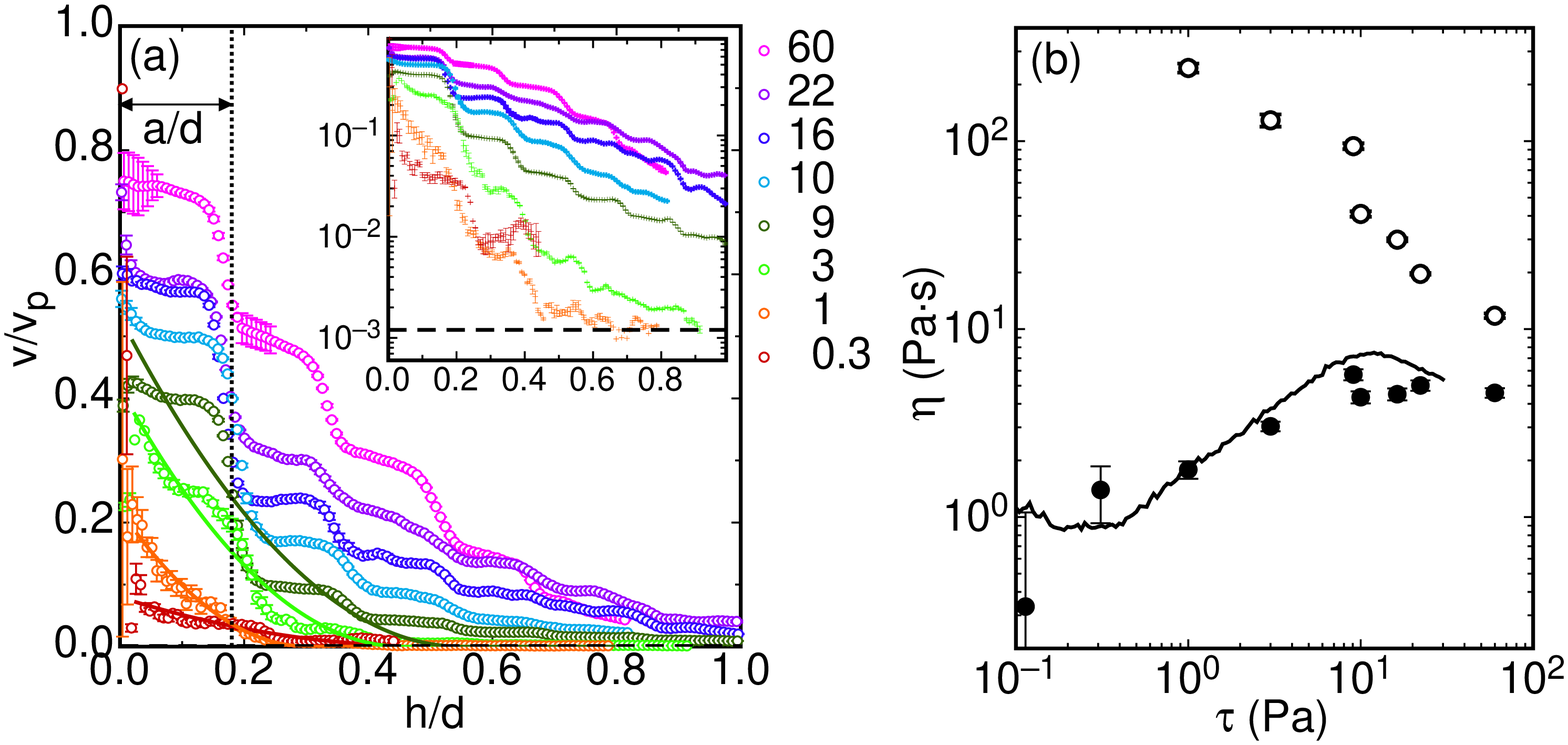}}
\caption{ (a) Shear profiles at the plate edge for settling particles of ZrO$_2$ in mineral oil ($\Delta\rho=2.9$ g/mL).  The mean velocity $v$ is normalized by the plate edge velocity $v_p$, and the depth $h$ is normalized by the gap $d$.  Shear stress $\tau$ for each profile shown in the key;  higher curves correspond to larger $\tau$.   Dashed line:  upper bound of $10^{-3}$ for a measurement at $\tau<\tau_{min}$.   Dotted black line:  depth equal to 1 particle diameter.  Solid lines: fits of Eq.~\ref{eqn:quadprofile} to the data for $\tau<\tau_{max}$.  Inset: same data on log-linear scale.  (b)  Local viscosity curves based on the local shear rate from the shear profile.  Open circles:  local viscosity in bulk region.  Solid circles: local viscosity in the shear band near the top plate.  Solid line:  global viscosity curve.  
}
\label{fig:shearprofileZrO}
\end{figure}
 
We first describe results for a settling suspension of 150 $\mu$m ZrO$_2$ spheres in mineral oil at $\phi=0.53$ with a gap $d=890$ $\mu$m.  This suspension is chosen for visualizations instead of glass because the particles are opaque.  The raw particle motions under shear are shown in supplementary videos 1 and 2 for two different shear rates.  Velocity profiles are shown in Fig.~\ref{fig:shearprofileZrO}a for a range of shear stresses.  The corresponding global viscosity curve is shown in Fig.~\ref{fig:shearprofileZrO}b.  Below $\tau_{min}\approx 0.3$ Pa, we found no measurable particle motion up to a resolution of $10^{-3}$ times the plate displacement.  In this regime the particles remained settled due to gravity as expected based on the measurements of $\tau_{min}$ in Sec.~\ref{sec:onsetstress}.  Above $\tau_{min}$, we found a narrow shear band near the moving top plate.  This also agrees well with the observations that the onset of shear thickening corresponds to the onset of dilation \citep{MW58}, since shearing of the grains is what results in dilation.  The width of the shear band increased as the stress was increased.  Layering was clearly observed at higher shear rates, which results in the step-like shear profiles in Fig.~\ref{fig:shearprofileZrO}a.   Effects of this layering on the measured stress are only expected for smaller gaps, less than about 5 layers \citep{BZFMBDJ10}.  We performed similar measurements with glass particles in mineral oil with a gap 12 particles wide as opposed to 6 particles with the ZrO$_2$.  Results were qualitatively similar to the ZrO$_2$ data, although layering was less prominent as expected for a finite-size effect, appearing only clearly in the top two layers.  


 

\begin{figure}
\centerline{\includegraphics[width=5.5in]{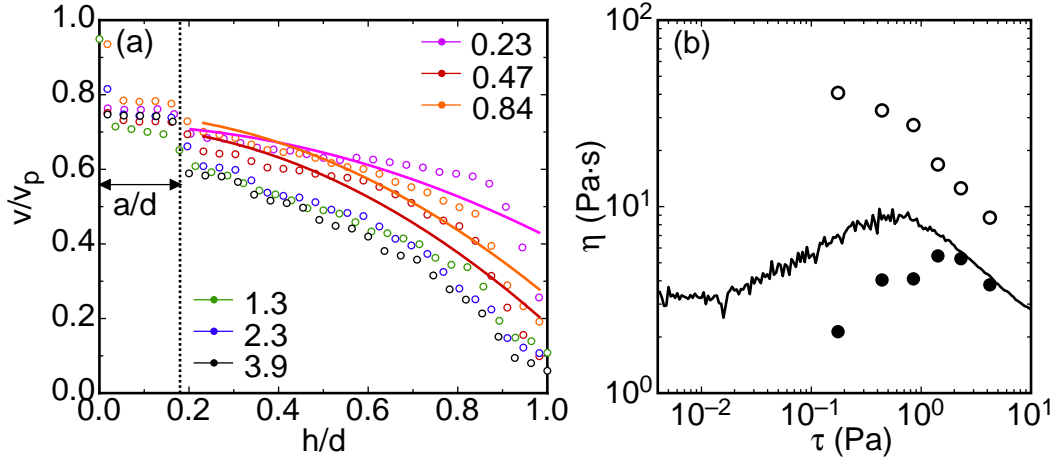}}
\caption{(a) Shear profiles at the plate edge for density matched polyethylene in silicone oil ($\Delta\rho \approx -0.01$ g/mL).  The mean velocity $v$ is normalized by the plate edge velocity $v_p$, and the depth $h$ is normalized by the gap $d$.  Shear stress $\tau$ for each profile shown in the key; lower curves correspond to larger $\tau$.  Dotted black line:  depth equal to 1 particle diameter.  Solid lines: fits of Eq.~\ref{eqn:quadprofilelin} to the data for $\tau<\tau_{max}$ with the substitution $h/d \rightarrow 1-h/d$ since the particles are lighter than the liquid.  (b)  Local viscosity curves based on the local shear rate from the shear profile.  Open circles:  local viscosity in bulk region.  Solid circles: local viscosity in the shear band near the bottom plate.    Solid line:  global viscosity curve.  
}
\label{fig:shearprofilepolyethylene}                                        
\end{figure}

We next describe results for nearly density-matched 135 $\mu$m polyethylene spheres in silicone oil at $\phi=0.55$ with a gap 850 $\mu$m wide.   The lighting used to take the videos heated the silicone oil by several degrees, so there was a slight density difference of about $\Delta\rho = -0.01$ g/mL such that the particles were slightly buoyant, effectively reversing the direction of gravity.   Because the smaller density difference moves the onset of shear thickening to very low shear rates, we did not obtain measurements of the shear profile below $\tau_{min}\approx 0.01$ Pa.  Velocity profiles are shown in Fig.~\ref{fig:shearprofilepolyethylene}a for a range of shear stresses.  The corresponding global viscosity curve is shown in Fig.~\ref{fig:shearprofilepolyethylene}b.  In the shear thickening regime, the velocity gradient in the bulk was relatively small, with a shear band at the bottom plate and a layered structure at the top plate. In this case, the shear band appeared at the bottom plate and the direction of curvature of the shear profile was reversed due to the inversion of gravity.  The shear band widened at higher stresses, similar to the case for the settling ZrO$_2$.  Interestingly, for all three suspensions the layering was most pronounced near the top moving plate despite the gravity inversion.  

These measurements also allow us to measure slip directly.  When the particles were settled with the plate moving past, there was not even contact so the difference between plate and particle motions is technically not slip and so would not be expected to follow slip correction models which usually assume a linear bulk velocity profile.  Settling and slip effects can be distinguished by comparing to the density matched case shown in Fig.~\ref{fig:shearprofilepolyethylene}a where the settling rate is much lower than the shear rate for all of the data shown.  The difference between the speed of the top plate and neighboring particles in the more developed flow regimes is around 25\%, roughly independent of shear stress.   This does not change significantly at the boundary between shear thickening and shear thinning regimes, confirming that those rheological boundaries are not determined by slip.  Since the goal of this paper is to understand the global response of Discontinuous Shear Thickening, we do not `correct' for slip.  Making a correction for slip would not significantly alter the shape of the viscosity curves nor move the regime boundaries in terms of stress because it only affects the shear rate, although it would slightly shift the magnitude of the viscosities reported.  The lack of contact between the particles and plate with settling is not problematic in terms of the mechanical response because viscous interactions within the liquid transmit stress between them just as well as hard contacts.  This is confirmed by our observation that switching from smooth to rough plates does not change the stress scales or whether Discontinuous Shear Thickening occurs.

\subsection{Local constitutive relations}

Here we use the shear profile to test constitutive relations in the shear thickening regime.  Since the shear stress in a parallel plate geometry is independent of height, a local hydrodynamic constitutive relation $\tau(\dot\gamma_l)$ dependent only on a local shear rate $\dot\gamma_l$ would correspond to a linear velocity profile.  To explain a non-linear steady state velocity profile, models have been introduced in the past to account for fluctuations in the local shear rate to an effective kinetic temperature \citep{NB94, BLSLG01} and the effect of the local variation in packing fraction on the viscosity \citep{NB94, BLSLG01, FLBBO10}.   In granular shear flows, the initial inhomogeneity is usually attributed to dilation near the moving plate \citep{MDKENJ00}.

Here, since the curvature of the shear profiles in Figs.~\ref{fig:shearprofileZrO} and \ref{fig:shearprofilepolyethylene} changed with the direction of effective gravity for the particles, we suggest that here gravity and friction are responsible for curvature in the shear profile, as in sedimenting flows \citep{LSP05}. Specifically, there can be frictional forces between particles due to the weight of the packing which increases with depth $h$ into the sample relative to the top plate (for downward gravity) if the particles remain in contact via force chains.  A non-linear shear profile could be the result of such an explicit height-dependence.  The simplest form for a local stress relation that includes gravity is

\be
\tau = \eta_{\nu}\dot\gamma_l + \tau_g h/d+\tau_c
\label{eqn:constitutivelaw}
\ee

\noindent where $\eta_{\nu}$ is the viscous hydrodynamic contribution to the viscosity, the gravitational stress scale $\tau_g \equiv \mu_{eff}\Delta\rho g d/15.3$ from Sec.~\ref{sec:onsetstress}, and $\tau_c$ represents any stresses that are independent of local shear rate and depth such as interparticle attractions.  Rearranging gives the local shear rate

\be
\dot\gamma_l = (\tau-\tau_c-\tau_g h/d)/\eta_{\nu} \ .
\label{eqn:shearrateloc}
\ee

\noindent This implies a critical depth $h_c/d = (\tau-\tau_c)/\tau_g$ at which the shear rate equals zero and beyond which there is no shearing of grains.  This suggests the shear stress must exceed the sum of gravitational stress and interparticle stresses (included in $\tau_c$) on the first layer of particles ($\tau_g a/d$) to shear grains, in agreement with the condition for the onset of shear thickening shown in Fig.~\ref{fig:sizestressgravity} and \citet{BFOZMBDJ10}, respectively.  The velocity profile can be obtained by integrating the local shear rate from Eq.~\ref{eqn:shearrateloc} over depth.  There are two solution regimes:\\
 if $h_c < d$, then 
 
\be
\frac{v}{v_p} =  \frac{\tau_g}{2\tau_{\nu}}\left(\frac{h_c-h}{d}\right)^2 \ ,
\label{eqn:quadprofile}
\ee

\noindent and if $h_c > d$, then

\be
\frac{v}{v_p} = \frac{\tau-\tau_c-\tau_g}{\tau_{\nu}}\left(\frac{d-h}{d}\right)+ \frac{\tau_g}{2\tau_{\nu}}\left(\frac{d-h}{d}\right)^2
\label{eqn:quadprofilelin}
\ee

\noindent where the plate velocity $v_p = d\dot\gamma$ and the viscous stress scale is defined by $\tau_{\nu} \equiv \eta_{\nu}\dot\gamma$. The curvature of the velocity profile characterized by a quadratic term is set by the ratio of gravitational to viscous stresses $\tau_g/\tau_{\nu}$.   The velocity profile becomes linear in the limit where this ratio goes to zero ($h_c >d$)  as expected.  These profiles are concave up, and become more linear with increasing $\tau$ in qualitative agreement with the data in Fig.~\ref{fig:shearprofileZrO}a.   The equations were written for the case where the effective gravity on the particles is downward.  For the polyethylene data where the effective gravity is upward, we have to make the substitution $h/d \rightarrow 1-h/d$ which reverses the concavity.

 \begin{figure}                                                
\centerline{\includegraphics[width=3in]{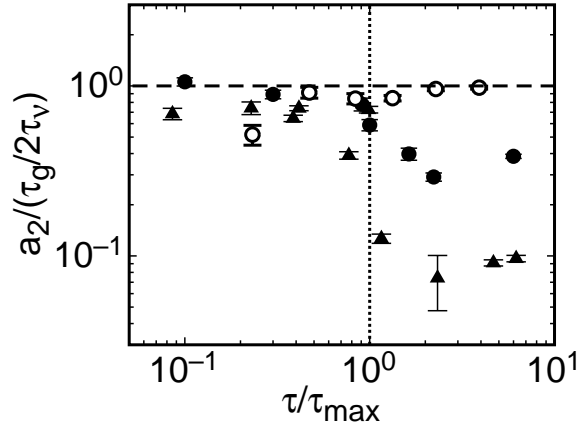}}
\caption{Quadratic curvature $a_2$ obtained from fit of Eqn.~\ref{eqn:profilefit} to velocity profiles.  The curvature is normalized by the model prediction from Eq.~\ref{eqn:quadprofilelin} with $\tau_{min}$ used as an estimate for the viscous stress $\tau_{\nu}$.  Data is fit for different normalized shear stresses $\tau/\tau_{max}$ for glass in mineral oil (solid triangles, $\Delta\rho = 1.58$ g/mL), polyethelyne in silicone oil (open circles, $\Delta\rho = -0.01$ g/mL), and ZrO$_2$ in mineral oil (solid circles, $\Delta\rho = 2.9$ g/mL).  The data collapse close to a value of 1 for $\tau < \tau_{max}$ suggests the curvature of the shear profile is due to the weight of the particles on deeper layers which is transferred via frictional contacts, and that the  contribution of viscous stresses to the viscosity does not increase significantly in the shear thickening regime. 
}  
\label{fig:profilequadfit}                                        
\end{figure}

Because we are applying a continuum model to a system that is quantized due to layering, and fluctuations could smooth out mean shear profiles, this model will only be able to crudely approximate the slope and curvature of the shear profile.  To test this model, we fit the function  

\be
v/v_p = a_1\frac{(h_c-h)}{d} + a_2\left[\frac{(h_c-h)}{d}\right]^2
\label{eqn:profilefit}
\ee

\noindent to the measured velocity profiles for each shear rate.  Some of these fits are shown in Figs.~\ref{fig:shearprofileZrO}a and \ref{fig:shearprofilepolyethylene}a.  For ZrO$_2$ and glass, it appears that $h_c < d$ for $\tau < \tau_{max}$ so we fix $a_1=0$ according to Eq.~\ref{eqn:quadprofile}.  The quadratic coefficient $a_2$ can be compared to the prediction of Eqns.~\ref{eqn:quadprofile} and \ref{eqn:quadprofilelin} with an estimate for the viscous stress $\tau_{\nu}$.   We showed in  Sec.~\ref{sec:onsetstress} that at the onset of shear thickening the viscous stress must be just enough to initiate shear so $\tau_{\nu} \approx \tau_{min}$.  Since the shear rate increases slowly in the shear thickening regime for Discontinuous Shear Thickening, we will use this estimate for the entire shear thickening regime.  The measured curvature $a_2$ normalized by the predicted value $\tau_g/2\tau_{min}$ is plotted in Fig.~\ref{fig:profilequadfit} for each fit for ZrO$_2$, polyethylene, and glass.  

In the shear thickening regime ($\tau < \tau_{max}$), the data for all three density differences collapse onto a single curve with $a_2/(\tau_g/2\tau_{min}) \approx 1$.  This value is in agreement with the model which confirms the curvature of the shear profile is set by a balance of gravity-induced friction and viscous interactions throughout the shear thickening regime.  This balance implies the weight of the packing builds up in deeper layers, which requires force chains of solid particle contacts must extend from plate to plate, which is a common feature of granular systems.  Since the shape of the shear profile depends on the specific force balance, the quadratic shear profile prediction is specific to settling suspensions within this model.  Because the data collapse works reasonably well for systems of different sizes, it rules out the role of finite size effects in setting the velocity profile curvature.  

The data collapse of the shear profile curvature to a constant value in the shear thickening regime also suggests that the contribution of viscous stresses to the global viscosity is not increasing significantly in the shear thickening regime, and remains close to $\tau_{min}$ , in contrast to the expectations of hydrodynamic models for shear thickening.  If the stress increase in the shear thickening regime was due to viscous forces proportional to shear rate, the reduced curvature would have to follow a slope of $-1$ in Fig.~\ref{fig:profilequadfit} in the range $\tau < \tau_{max}$.  Rather, the increase in stress in the shear thickening regime must be hidden in the uniform term $\tau_c$ due to other so-far-unidentified forces in Eqn.~\ref{eqn:constitutivelaw}.


For $\tau>\tau_{max}$, the different curvature values do not collapse onto a single curve, suggesting either that the model fails in this regime or at least that $\tau_{min}$ is no longer a good approximation of $\tau_{\nu}$ in this regime.  Rather, the profiles appear to be closer to exponential  (see inset of Fig.~\ref{fig:shearprofileZrO}a), similar to granular shear profiles of spherical particles \citep{MDKENJ00}.  This suggests that above $\tau_{max}$ the shear profile could be that of a fully granular system where there is no need for a contribution of viscous hydrodynamics.  


Curvature in the shear profile has been attributed to variations in the local packing fraction in some other experiments\citep{FLBBO10}.  In a hydrodynamic model, small changes in packing fraction from dilation and viscous resuspension are significant because of the viscosity divergence with packing fraction as the viscous lubrication layer goes to zero at the jamming transition.   However, our results on the curvature of the shear profile suggest stress is transmitted mostly through frictional contacts rather than viscous interactions.  In granular mechanics, frictional contact forces can vary by less than about 30\% with changes in packing fraction \citep{LW69}, and this contribution would be small in comparison to the separation of local viscosities seen in Figs.~\ref{fig:shearprofileZrO}b and \ref{fig:shearprofilepolyethylene}b and the variations in curvature in Fig.~\ref{fig:profilequadfit}.  Another reason to suspect that viscous forces would not account for the measured stresses in the shear thickening regime comes from the magnitude of the measured viscosity.  Since viscosity values measured are up to $10^7$ times the solvent viscosity \citep{BJ09}, this would require subatomic gaps in a lubrication model. However, lubrication in molecular liquids breaks down at 2 molecular layers, below which the liquid is frictional \citep{VG88}.  These issues suggest the stresses in the shear thickening regime must be explained by some non-viscous mechanism.

Another model attributes curvature in the shear profile to a gradient in kinetic energy due to fluctuations in particle motion \citep{NB94, BLSLG01}.  The contribution of this effect to the local stress gradient can be estimated as $\nabla \tau \sim \rho v\nabla v \sim \rho \dot\gamma^2 d$ using the observation that the scale of the rms fluctuations in velocity are comparable to the mean flow velocity in granular flows with solid particle contacts \citep{BLSLG01}.  This contribution to the stress gradient is at most on the order of $10^{-4}$ times the gravitational contribution to the stress gradient $\nabla \tau \sim \Delta \rho g$ even at the maximum stress in the shear thickening regime for the measurements shown in Figs.~\ref{fig:shearprofileZrO}a and \ref{fig:shearprofilepolyethylene}a.  Thus we expect the contribution of the kinetic energy to the shear profile to be negligible in the regime of our measurements. 


 \subsection{Localized viscosity curves}


Here we investigate the validity of local constitutive relations by plotting local viscosity curves.  The local viscosity can be calculated as the ratio of the measured global stress and the local shear rate from the derivative of the velocity profile.  To separate the bulk region from the shear band regions we use the mean slope of the velocity profile over different ranges of depth.  For the profiles in Fig.~\ref{fig:shearprofileZrO}a we use the range $0.25 < h/d < 0.8$ for the bulk, and $h/d <0.3$ for the shear band at the top plate.   For the profiles in Fig.~\ref{fig:shearprofilepolyethylene}a we use the range $0.2 < h/d < 0.8$ for the bulk, and $h/d > 0.85$ for the shear band at the bottom plate.  According to the model of Eqns.~\ref{eqn:quadprofile} and\ref{eqn:quadprofilelin}, the local bulk viscosities roughly approximate the linear term, which corresponds to the viscous contribution to the viscosity.  These local viscosities are plotted in Fig.~\ref{fig:shearprofileZrO}b and \ref{fig:shearprofilepolyethylene}b, respectively, along with the global viscosity curves.  For each suspension, the shear band shows shear thickening similar to the global curve, while the bulk region appears to be everywhere shear thinning based on the local viscosity.  In the non-density matched case below $\tau_{min}$, the bulk was observed to be settled (i.e. locally jammed), corresponding to an infinite local viscosity.  Thus the bulk region appears to have a yield stress due to gravity and consequently shear thinning but no shear thickening.  The region that qualitatively determines the global rheology is not the bulk but rather the near-wall region where the shearing occurs. This is not surprising from a granular or solid mechanics point of view where the global behavior is often determined by the failure in the weakest region.  



Another test of the validity of local constitutive equations comes from comparisons of measurements in different measuring geometries.  For a measurement in a cone and plate geometry the mean shear rate is independent of radius because the plate speed is proportional to the gap height at each point along a radius, while in a parallel plate setup the mean shear rate increases with radius because the plate speed is faster near the edge but the gap remains the same.  Thus, assuming a local constitutive relation between shear stress and rate holds, an apparent viscosity curve measured in a parallel plate rheometer should always be smoothed out compared to one measured in a cone and plate rheometer.  However, a comparison of measurements of Discontinuous Shear Thickening in different geometries showed that the apparent viscosity curve from a parallel plate rheometer was {\em steeper} than that measured with a cone and plate rheometer \citep{FHBOB08}. This discrepancy between measuring geometries indicates that the constitutive relation for the stress is not a function only of the local shear rate, even for that density matched suspension.  

In this section we tested a constitutive relation that attributes the shape of the shear profile in the shear thickening regime for  settling suspensions to the balance between viscous forces and gravitational forces for particles in frictional contact.  The onset stress $\tau_{min}$ occurs at the onset of particulate shear.  However, the remarkable feature of Discontinuous Shear Thickening is the large stress jump in the shear thickening regime.  We found that local constitutive relations between stress and shear rate fail to describe the stress in the shear thickening regime.  Specifically, the collapse of the shear profile curvature values in Fig.~\ref{fig:profilequadfit} for $\tau<\tau_{max}$ suggests the viscous term proportional to local shear rate is nearly constant in the shear thickening regime, and the stress jump can not be attributed to viscous forces or the inhomogeneity due to gravity.  Instead, the stress jump must be hidden within the additional stress term $\tau_c$ from Eqn.~\ref{eqn:constitutivelaw}, whose source will be identified in Sec.~\ref{sec:normalstress}.   


\section{Normal forces and the boundary condition}
\label{sec:normalstress}


In this section we use measurements of shear and normal stresses under different boundary conditions to show that the global mechanical response can be described by a solid frictional constitutive law rather than  a viscous law.  

\subsection{Frictional scaling}

\begin{figure}                                                
\centerline{\includegraphics[width=6.5in]{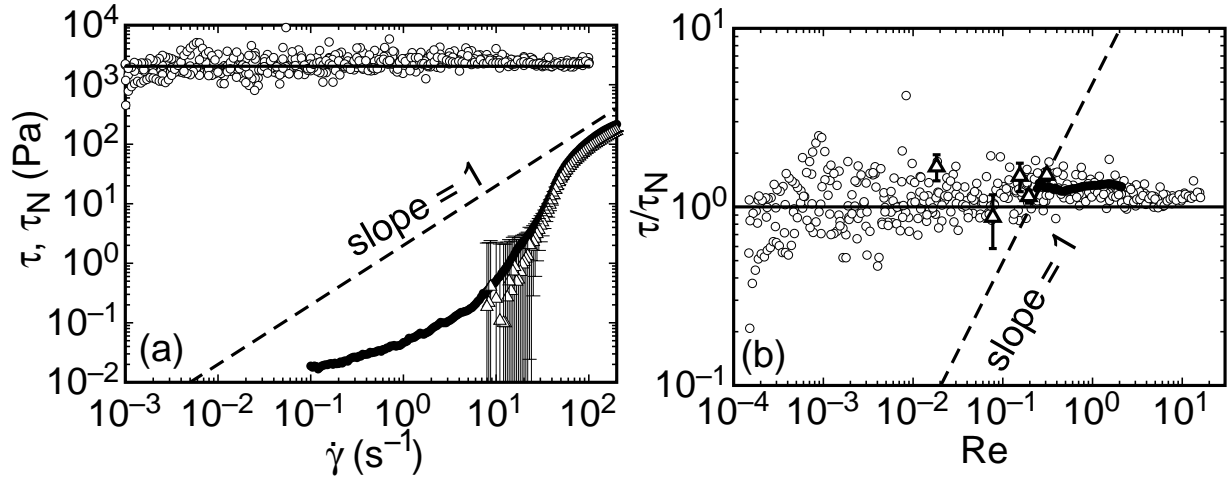}}
\caption{(a) A comparison of flow curves measured with different boundary conditions for glass spheres in water at $\phi=0.52$ ($<\phi_c$).  Solid circles: shear stress $\tau$ for 100 $\mu$m spheres in a fixed gap measurement with the standard parallel plate setup.  Open triangles:  normal stress $\tau_N$ from the same measurement.  The absolute uncertainty on the normal stress is 2 Pa, so the normal stress cannot be resolved at the low end.  Open circles: $\tau$ for 500 $\mu$m spheres with a fixed normal stress of 2040 Pa (solid line) in the modified parallel plate setup with a hard wall.  Dashed line:  slope 1 corresponding to a Newtonian scaling for reference.  (b)  Circles: same data with the shear stress $\tau$ normalized by normal stress $\tau_N$ vs. $Re=\rho d^2\dot\gamma/\eta_m$.  Open triangles:  constant shear rate measurements in the standard parallel plate setup in which the normal force was recalibrated before each measurement.  Solid line:  $\tau= \tau_N$  indicating a frictional scaling.  Dashed line:  $\tau=\eta_m\dot\gamma$ corresponding to a viscous scaling.  Dotted line:  normal stress resolution limit.
}
\label{fig:normalstress}                                        
\end{figure}

Here we compare steady state viscosity curves along with normal stress measurements for similar suspensions with different boundary conditions.  The sample was 100 $\mu$m diameter glass spheres in water at a packing fraction of $\phi=0.52$ ($<\phi_c$).   We first show results from a standard parallel plate setup (Fig.~\ref{fig:rheometer}a) with a diameter of 50 mm which results in a better normal stress resolution than smaller plates.  The shear stress $\tau$ and normal stress $\tau_N$ are shown in Fig.~\ref{fig:normalstress}a as functions of shear rate $\dot\gamma$ for a measurement in which the gap size is fixed at 0.72 mm.  The region with slope greater than 1 defines the shear thickening regime.  We found positive normal stresses, corresponding to the sample pushing against the plates, in agreement with other measurements of Discontinuous Shear Thickening \citep{LDHH05, FHBOB08}.  The shear and normal stresses track each other extremely well in functional form and magnitude.   The cutoff of $\tau_N$ at the low end corresponds to the measurement dropping below the relative resolution of the normal stress of about 0.3 Pa. 

We next used the walled rheometer setup without a liquid air interface or room for expansion as shown in Fig.~\ref{fig:rheometer}b.  In this case a sample of 500 $\mu$m glass spheres in water was used; the larger particles were necessary to avoid them escaping through the gap between the side wall and top plate.  While the values of $\tau_{min}$ and $\tau_{max}$ differ with particle size (see Figs.~\ref{fig:sizestressgravity}, \ref{fig:stressmaxcollection}), otherwise the samples behave in a qualitatively similar way in the normal parallel plate setup.  The normal force on the top plate was fixed at 1 N (2040 Pa), consequently the gap size was allowed to vary.  The viscosity curve is shown in Fig.~\ref{fig:normalstress}a.  In contrast to the standard parallel plate setup, the rheology is that of a yield stress fluid with no shear thickening regime.  Such a dramatic difference in behavior with a change in boundary conditions would be unexpected from a local hydrodynamic constitutive relation, and implies a non-local effect.  The common feature of both measurements is the connection of the shear stress to the normal stress.  We plot the ratio of stresses $\tau/\tau_N$ vs.~the Reynolds number $Re = \rho_l d^2\dot\gamma/\eta_m$ for both measurements in Fig.~\ref{fig:normalstress}b.  Additionally, we show steady state values for measurements taken at constant shear rate in the shear thickening regime in which the normal force was recalibrated relative to the static value before each measurement to optimize resolution of the relative normal force.  The fact that these three data sets under different measurement conditions collapse onto the same curve suggests a global constitutive relation independent of boundary conditions.   Since $\tau/\tau_N$ is near unity and constant over five decades of Reynolds number this suggests that the measured stresses are compressional in nature.  A compressional scaling from either viscous, inertial, or frictional forces in dense suspensions is the result of the redirection of stress in different directions through the bulk of the suspension by particle interactions \citep{NB94, PK95, BV95, JNB96, SB02, DGMYM09}.   This compressive scaling can also account for the constant stress term $\tau_c$ in the constitutive relation of Eq.~\ref{eqn:constitutivelaw}, since it is independent of local shear rate and height, but instead is dependent on the normal stress.  A compressional scaling also confirms the total normal stress is the relevant physical quantity, rather than normal stress differences.

 While viscous or inertial stresses can in some cases result in constant $\tau/\tau_N$, the stresses must be proportional to shear rate or shear rate squared, respectively, neither of which is satisfied for Discontinuous Shear Thickening suspensions as in Fig.~\ref{fig:normalstress}.   Rather, a frictional explanation in which the forces are transmitted along chains of hard particles via frictional contact \citep{JNB96} is supported by the observations that the stresses have no inherent shear rate dependence and continue to follow this same relation when additional normal stress is applied at the boundary as shown in Fig.~\ref{fig:normalstress}, the indication from the shear profile that the weight of the packing is transmitted along frictional contacts to build up in lower layers shown in Sec.~\ref{sec:shearprofile}, and the relative smallness of the expected viscous and inertial forces in the Discontinuous Shear Thickening regime, as explained in Sec.~\ref{sec:inertial}.  



\subsection{Transient normal force control measurements}

Here we show that the coupling between shear and normal stress applies even to transient measurements as the normal force boundary condition changes, and that shear thickening can be eliminated if the normal force is removed from the boundary as suggested by \citet{FHBOB08}.  To emphasize the generality of these results, we show this result for a different suspension, cornstarch in water.   Both this result and the results of the previous section were found for both suspensions, but we show only one set of data for brevity.

\begin{figure*}                                                
\centerline{\includegraphics[width=6.5in]{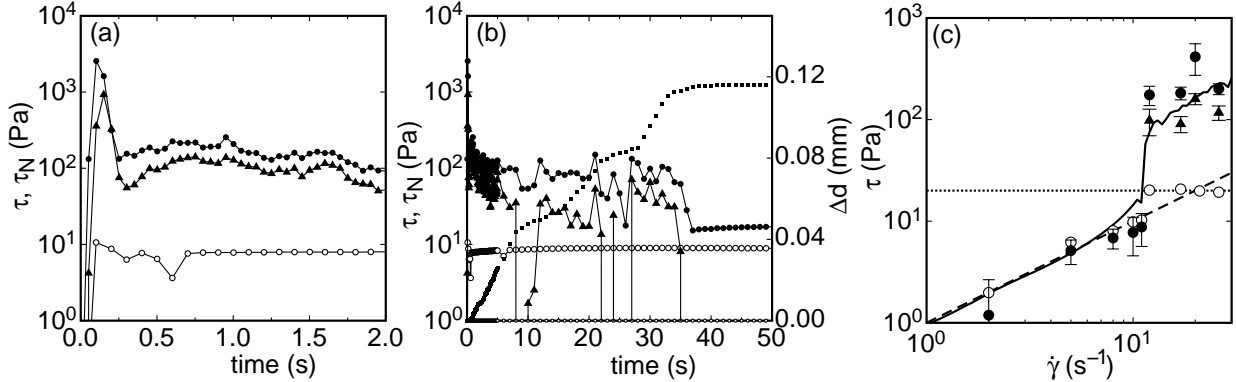}}
\caption{(a) Transient time series of shear stress $\tau$ (circles) and normal stress $\tau_N$ (triangles) in normal force control measurements for cornstarch in water at $\phi=0.55 < \phi_c$.  The sample starts at rest then the shear is switched on at time $t=0$.  Solid symbols: shear rate $\dot\gamma= 26$ s$^{-1}$ (above $\dot\gamma_c$).  Open symbols: $\dot\gamma=8$ s$^{-1}$ (below $\dot\gamma_c$).  (b)  Same data as panel a, but extended to longer times to see the steady state behavior. Right axis: change in gap size $\Delta d$ (small squares).   (c) Effective viscosity curves obtained from transient measurements.  Solid circles:   transient shear stress averaged between 0.4 and 1.0 s after shear starts. Solid triangles: transient normal stress averaged over the same time.  Open circles: steady state shear stress at the end of the time series where $\tau_N$ was below the resolution limit for each shear rate.   Discontinuous Shear Thickening is suppressed when the normal stress at the boundary is removed.  Solid line:  stress as a function of shear rate obtained from a steady state viscosity curve for the same sample with fixed gap size.  Dashed line: Newtonian scaling.}
\label{fig:viscnormalforcecontrol}                                        
\end{figure*} 

We performed normal force controlled experiments modeled after those of \citet{FHBOB08}.  These measurements were done in the standard parallel plate setup in a normal force controlled mode. The normal force set point is zero relative to the rest state, with an initial gap of $d=1.08$ mm.  The gap size is free to vary during the measurements to adjust the normal force back to the setpoint via a feedback loop.  Initially the sample of cornstarch in water at $\phi=0.55 < \phi_c$ was at rest, then at time $t=0$ the shear rate was set to a constant non-zero value for the rest of the experiment.  Examples of transient time series of the shear stress and normal stress are plotted in Fig.~\ref{fig:viscnormalforcecontrol}a for two different shear rates.  For shear rates below the onset of shear thickening $\dot\gamma_c \approx 11$ s$^{-1}$, the stress quickly came to near the steady state value within a fraction of a second and remained there.  For shear rates above $\dot\gamma_c$, the shear and normal stresses had a large peak initially, exceeding the steady state value by more than an order of magnitude.  Even though the normal force set point was zero, the normal stress can be non-zero in the transient behavior as the gap adjusts via a feedback loop.  Longer time series are shown in Fig.~\ref{fig:viscnormalforcecontrol}b along with the variation in gap size.  Below $\dot\gamma_c$ the normal force did not exceed the threshold to cause the gap to move.  In contrast, above $\dot\gamma_c$ the gap increased initially due to the transient normal force.  The shear stress tracked the normal stress quite well throughout the entire transient process,  and they were similar in magnitude.  The stresses each decreased as the gap increased, and the gap stopped increasing when the normal stress dropped below the feedback threshold of 20 Pa.  Beyond this point the stresses and gap size remained constant, which was measured for at least 200 s in each experiment to confirm that the system was in a steady state.  

We summarize the normal force control experiments with effective viscosity curves in Fig.~\ref{fig:viscnormalforcecontrol}c.  We show the transient shear and normal stresses averaged between 0.4 and 1 s after the shear was started as solid symbols.  Because the response time of these samples to dramatic changes is typically a fraction of a second, while the normal force control feedback loop has a longer timescale, these transient results effectively correspond to a fixed gap boundary condition.   They show the same qualitative shear thickening as steady state behavior for fixed gap measurements, indicated by the solid line.  Differences between the solid circles and solid line beyond the measurement resolution indicate a difference between steady state and transient measurements.  Stress values taken from the end of the test, where $\tau_N=0$ ($\pm 20$ Pa) and the system was in a steady state, are shown as open circles in Fig.~\ref{fig:viscnormalforcecontrol}c.  The effective viscosity curve based on this data is consistent with a Newtonian scaling at shear rates below $\dot\gamma_c$.  Above $\dot\gamma_c$, the shear stress values match up with the normal stress feedback threshold.  This can be understood since a normal stress of that magnitude is not enough to trigger the normal force control feedback loop so the gap size remained fixed, but the normal stress can still couple to the shear stress.  The strong shear thickening in the fixed gap and transient data is totally absent from the $\tau_N=0$ data.  We note that there is no significant dependence of viscosity curves on gap size in this range \citep{BZFMBDJ10}, so the difference must be due to the fact that the normal force is fixed to be zero.  This shows that a positive normal stress of comparable magnitude is required to achieve the shear stress associated with shear thickening.  In the absence of this confining stress, shear thickening cannot occur, as was suggested by \citet{FHBOB08}.   Without making any assumption about the mechanism for coupling between the normal and shear stresses at values below the normal stress feedback threshold, the open circles in Fig.~\ref{fig:viscnormalforcecontrol}c put an upper bound on the viscous and other non-compressive contributions to suspension viscosity and show that they are not responsible for  Discontinuous Shear Thickening.  This directly shows the surprising result that cornstarch in water, the prototypical Discontinuous Shear Thickening suspension, is not actually shear thickening based on the direct constitutive relation between stress and shear rate when boundary conditions are held constant, and the occurrence of the phenomenon depends on a changing contribution of normal stresses from the boundary.

We have noted that the normal and shear stresses track each other quite well in normal-force-controlled measurements.  In fact, in all of the various types of experiments on suspensions that exhibit Discontinuous Shear Thickening in which normal stresses and shear stresses were compared they tended to track each other quite well.   For example, we attempted measurements with a fixed normal stress $\tau_N=0$ and fixed shear stress greater than the normal stress feedback threshold.  Since the shear stress is the dominant control parameter of the rheometer, the shear stress reached the set value but the measurements never reached a steady state because the normal stress could not drop, causing the gap to increase until the top plate detached from the sample.  Similarly, in experiments by \citet{LDHH05} that measured stress fluctuations in the steady state, fluctuations of the normal stress and shear stress were found to be strongly coupled with a proportionality close to 1.  This helps explain an earlier result in which an apparent viscosity curve no longer showed shear thickening when positive fluctuations in the shear stress in the steady state were removed from the data \citep{LDH03}.  Since the shear stress fluctuations were associated with the normal stress, this was in essence showing an effective viscosity curve with no normal stress.  Another example comes from our measurements of finite-size effects at very small gap sizes \citep{BZFMBDJ10}, in which the normal stress scaled with and was close in magnitude to the shear stress as it varied with gap size.  

In this section, we showed that the shear and normal stresses are coupled with a proportionality coefficient close to 1 in the shear thickening regime, and the relationship between stress and shear rate even changed with the boundary conditions to satisfy this stress coupling.  These results suggest that the stresses responsible for Discontinuous Shear Thickening are transmitted by frictional compressive forces.

\section{Dilation}
\label{sec:dilation}  
 
 \begin{figure}                                                
\centerline{\includegraphics[width=3.75in]{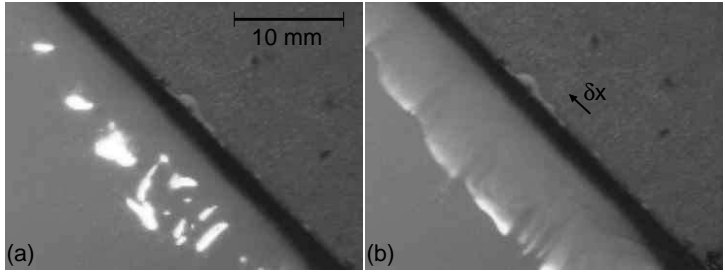}}
\caption{Top views of a 2.4 mm deep layer of cornstarch in water.  (a): Below $\phi_c$ in a shear cell at rest.  (a) at a shear rate above $\dot\gamma_c$, taken after a shear displacement of $\delta x = 2.5$ mm relative to panel a.  Dilation can be observed as an increase in surface roughness in the sheared region near the wall.
 }

\label{fig:dilationimages}                                        
\end{figure} 
 
The previous section showed that the normal stress boundary conditions controls the shear stress response.  This implies that the normal stress must at the boundary must increase dramatically with shear under typical measurement conditions that produce Discontinuous Shear Thickening.  The compressional scaling implies stresses are redirected through the system in all directions, so that similar confining pressures must be maintained on the system on all sides -- even a suspension-air interface such as at the side of a parallel plate rheometer.  In this section we analyze images of the suspension surface in contact with air to identify the boundary conditions responsible for the normal stress.  In systems that show Discontinuous Shear Thickening, it has been long known that there can be a visible change in the surface of suspensions at the onset of shear thickening \citep{MW58, OM00, SBCB10}.  It was understood early on that this was due to dilation, which occurs along with normal stresses \citep{MW58, Re1885}.  When wet granular packings dilate under shear, they take up more space than at rest, and consequently the liquid is then sucked away from the boundary into the enlarged interstices between grains, so by eye the surface appears to become rough as the particles poke through.  

This visible effect of dilation is shown for a suspension of cornstarch (14 $\mu$m) in water below $\phi_c$ in Fig.~\ref{fig:dilationimages}.  The suspension was in a 2.4 mm deep layer and viewed from above.  One of the side walls could be displaced to shear the suspension.  Before shear the surface of the suspension looked wet and shiny, as seen in panel a.  When the upper right wall was sheared at a rate exceeding the onset of shear thickening the nearby suspension appeared rough, shown in panel b.  As soon as the shear rate dropped, the surface appeared smooth and shiny again.  This behavior is shown in supplementary video 3.  We observed that the onset of visible dilatancy corresponds closely to the onset of shear thickening, consistent with previous observations that the shear thickening corresponds to the onset of dilation \citep{MW58, SBCB10}. 

Usually we find suspensions will show Discontinuous Shear Thickening in rheological measurements if the surface changes from shiny to rough when sheared, indicating dilation.  At low packing fractions, the surface remains smooth under shear because the packing fraction is too low for granular dilation to affect the surface, since volume changes from dilation are typically only a few percent \citep{Re1885, OL90}.  Alternatively, if the suspension has a yield stress, the surface may be always rough and not change with shear rate, even if the packing still dilates with shear.  Thus, the conditions where a change in the surface from dilation is observed seem to correspond to the conditions for suspensions to show Discontinuous Shear Thickening.  

There is a notable exception to the rule that a visible change in the surface from dilation indicates shear thickening.  Settling particles in a Couette cell were seen to dilate but did not shear thicken, and instead a yield stress was measured \citep{MW58, FBOB09}.  However, the inhomogeneity due to gravity can explain this behavior.  The weight of the particles in a vertical column of height $H$ pushes on the movable side walls, resulting in a yield stress on the scale of $\Delta\rho g H$ which can be on the order of kPa for typical rheometer Couette cells \citep{FBOB09}, well above the shear thickening stress regime which we observed for glass spheres from 10-100 $\mu$m in a parallel plate geometry which does not measure this yield stress (see Fig.~\ref{fig:stressmaxcollection}).  In the case of vertical walls, the side of the suspension is still jammed at rest which prevents shear thickening, but the top is not, so the suspension falsely appears unjammed when viewed from the top.  A more general conclusion that applies regardless of inhomogeneities is that dilation is necessary for Discontinuous Shear Thickening but not sufficient because shear thinning stresses must be small compared to shear thickening stresses or else the shear thickening will be hidden \citep{BFOZMBDJ10}.   

\begin{figure}                                                
\centerline{\includegraphics[width=5.5in]{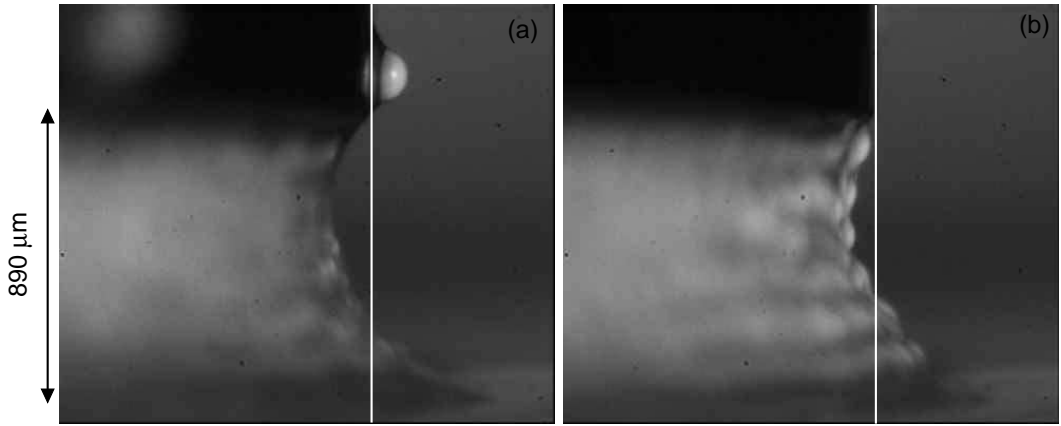}}
\caption{Images a suspension of 150 $\mu$m ZrO$_2$ particles in mineral oil in the standard parallel plate setup with a gap of  890 $\mu$m.  The camera is focused at a point on the edge of the suspension, with the line of sight tangent to the surface to view radial variations in the boundary position.  The rest of the image is out of focus because of the large amount of depth in the image.  (a) The suspension at rest.  (b) The suspension is sheared at constant shear rate of 3 Hz corresponding to $\tau_{max}$.  It can be seen that shear results in both radial dilation of the suspension and increased local curvature at the surface on the particle scale.  Vertical lines: reference lines indicating the plate edge in each image. 
}
\label{fig:dilationtangentview}                                        
\end{figure}

Now that we have established the importance of the boundary conditions, we want to directly address what the boundary looks like on the particle scale.  To this end we use a sample of opaque 150 $\mu$m diameter ZrO$_2$ particles in mineral oil at $\phi=0.54$ in a standard parallel plate rheometer setup with a gap size of 890 $\mu$m.  A video camera was focused at a point on the surface of the suspension with the line of sight tangent to the surface to best view any radial variations in the boundary position due to dilation.  The sample is shown at rest in Fig.~\ref{fig:dilationtangentview}a, in which case it had a smooth surface.   The same sample is shown in panel b at a steady shear rate of 3 $s^{-1}$, corresponding to $\tau_{max}$ at the upper bound of the shear thickening regime.  The boundary appears bumpy as particles penetrate the liquid-air interface.   The dynamic behavior is shown in supplementary video 4.   By reference to the edge of the rheometer plate (red line), it can be seen that the sample has expanded radially relative to the rest state by about 50 $\mu$m, corresponding to $0.3a$ or a volume increase of 0.8\%.   The penetration of the liquid-air interface by the particles can also be seen in the shear profile videos of ZrO$_2$ from Sec.~\ref{sec:shearprofile} as texture differences indicating contact lines on particle surfaces in supplementary videos 1 and 2.  

As long as the particles are between about 1 and 100 $\mu$m, then the grains are large enough to scatter light diffusively and small enough that they cannot be seen individually, so the surface appears rough by eye.  However, we note that for colloidal particles smaller than around 1 $\mu$m, the roughness of the surface becomes smaller than the wavelength of light and no longer scatters diffusively, so the surface of a colloid may even appear shiny if the surface is deformed by particles, and dilation would not be clearly visible.  This is confirmed, for example, by observations of stable asperities in a jammed colloid of 1.6 $\mu m$ diameter particles -- indicating stresses from surface tension -- in which the surface remained shiny \citep{KW11}.


In this section we suggested that the often-observed but unexplained connection between dilation and shear thickening is that dilation causes suspensions to interact with their boundaries and cause a change in boundary conditions.  This could explain the observation that the boundary conditions determine the constitutive relation between stress and shear rate and even whether Discontinuous Shear Thickening is observed.

\section{Capillary forces}
\label{sec:surfacetension}

In the previous section showed that when a dense suspension dilates under shear, the particles penetrate the liquid-air interface.  Here we propose a model by which dilations changes the boundary condition to produce a confining stress from capillary forces, which can provide the normal stress required for Discontinuous Shear Thickening.  Indeed, it has been suggested that capillary forces at boundaries could play an important role in the rheology of shear thickening suspensions \citep{HFC03,HCFS05, CAS05}. 

Changes in surface roughness similar to those from dilation have been observed in jammed suspensions \citep{CHH05}, and in free-surface flows, becoming more apparent at higher applied stresses and at higher packing fractions \citep{LNS02, TM05, SNS06}.  It was argued that these deformations required normal stresses in the suspensions to balance forces from surface tension due to the curvature of the liquid-air interface.  When dilation causes particles to penetrate the edge of the suspension to create a curved liquid-air interface, the scale of the radius of curvature $r$ of the liquid-air interface with surface tension $\gamma$ becomes comparable to the particle size $a$ \citep{LNS02}.  This produces a stress from surface tension pushing on the particles towards the interior of the suspension.  We estimate this stress from surface tension to be on the scale of $\gamma/r \sim \gamma/a \sim 100$ Pa for 100 $\mu$m particles, a significant stress in the context of the rheological measurements.  If the particles did not interact along stiff force chains, they would be pushed to the interior of the sample by this stress (assuming the liquid wets the particles, which is also a requirement to observe shear thickening \citep{BFOZMBDJ10}).  However, the fact that the particles continue to penetrate the surface in the steady state implies that, in the absence of inertial effects (satisfied by the low $Re$ of the experiments), forces must be transmitted all the way through the packing to balance the stress from surface tension.  The stress from surface tension can then be considered a confining stress which is transmitted and redirected through the suspension to the rheometer plates according to the compressional stress relation seen in Figs.~\ref{fig:normalstress} and \ref{fig:viscnormalforcecontrol}.  This stress would be the contribution that increases rapidly with shear rate as the system dilates under shear which is the characteristic feature of Discontinuous Shear Thickening.  Eventually, this confining stress reaches the limiting scale of $\gamma/a$ from surface tension, and beyond that point any additional shear stress must come from other sources, which are likely weak compared to the confining stress if shear thickening is observed, so the viscosity will likely drop off beyond the maximum confining stress.  Thus the maximum confining stress should correspond to $\tau_{max}$.  Above $\phi_c$ where the suspension is jammed, particles are seen penetrate the surface even without shear \citep{BZFMBDJ11}, so the yield stress scale $\tau_j$  should also be set by the confining stress from surface tension.  In the remainder of this section we quantitatively compare the measured shear stresses to the confining stress scale from surface tension $\gamma/a$ to test this model. 

We now address why surface tension determines the values of $\tau_{max}$ and $\tau_j$ rather than $\tau_{min}$.  We already argued that the confining stress from surface tension produces normal stresses between particles and the walls that results in a shear stress via friction.  This confining stress increases precipitously in the shear thickening regime as the amount of dilation increases with shear rate.  The maximum confining stress from surface tension should be on the order of $\gamma/a$ if $\delta$ reaches the order of $a$ or larger.  Beyond the point where the confining stress reaches its maximum, any additional shear stress must come from other sources, which are likely weak compared to the confining stress if shear thickening is observed, so the viscosity will likely drop off beyond the maximum confining stress.  Thus the maximum confining stress should correspond to $\tau_{max}$.  Above $\phi_c$ where the suspension is jammed, particles are seen penetrate the surface even without shear \citep{BZFMBDJ11}, so the yield stress scale $\tau_j$  should also be set by the confining stress from surface tension.


\subsection{Relating dilation to confining stress}

\begin{figure}                                                
\centerline{\includegraphics[width=3.in]{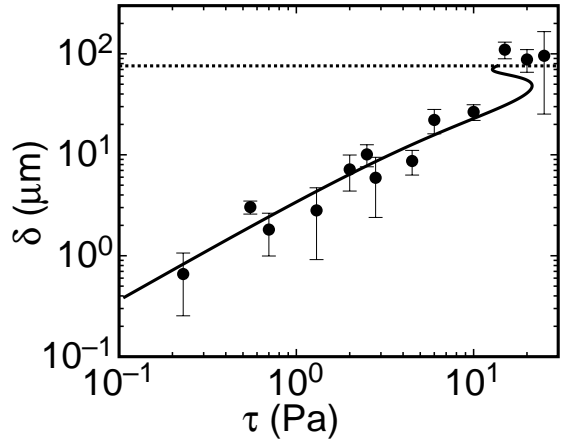}}
\caption{The radial dilation $\delta$ measured as a function of shear stress $\tau$ for 135 $\mu$m polyethylene spheres in silicone oil.  Solid line:  predicted relationship between $\delta$ and $\tau$ for a model in which there is a confining stress from surface tension $\tau \propto \gamma/r(\delta)$ where the local radius of curvature $r(\delta)$ is calculated geometrically.  A proportionality coefficient of 0.14 shifting the curve horizontally is used to fit the data. Dashed line:  dilation value where the contact line is expected to reach the 2nd layer of particles resulting in a dramatic increase in confining stress with dilation.
} 
\label{fig:dilationstress}                                        
\end{figure}

To quantify dilation, we measured the mean radial displacement of the surface during shear seen in a tangent view as shown in Fig.~\ref{fig:dilationtangentview}.  We did this for several different steady state shear rates in a sample of opaque 135$\mu$m polyethylene spheres in silicone oil at $\phi=0.56$.  For each measurement, we started the sample at rest, then sheared at a constant shear rate until the stress reached a steady state for some time, then stopped the shear to observe the relaxation to rest.  We repeated the cycle of shear followed by resting a total of 5 times.  The edge of the sample was tracked throughout these measurements.  The dilation $\delta$ was measured as the mean radial displacement of the edge between the steady state shear and rest states, averaging over the height of the sample and over at a period of at least 10 s and a strain of at least 2 in the steady state for each experiment.  The measured dilation is plotted versus the corresponding steady state stress values in Fig.~\ref{fig:dilationstress}.   Plotted error bars correspond to the standard deviation of $\delta$ measured over the 5 cycles.   All points shown correspond to stresses above $\tau_{min}$.  At lower stresses we could not resolve any dilation below our resolution limit of 0.5 $\mu$m.  The upper end of the shear thickening regime corresponds to $\tau_{max} = 2$ Pa for this sample.  

Using a geometric model, we can calculate a typical radius of curvature of the liquid-air interface as it contracts for a given particle dilation $\delta$, given the contact angle and conservation of liquid volume.   This allows us to estimate a confining stress scale from surface tension $\gamma/r(\delta)$.  Details of this calculation are shown in the appendix.  Briefly, the initial state with a relatively large radius of curvature corresponds to $\delta=0$.  As $\delta$ increases, the surface becomes curved as particles penetrate the surface due to dilation, and the radius of curvature decreases.   The corresponding stress scales almost linearly with $\delta$, and the scale of the radius of curvature is set by the particle size when the dilation is around a particle radius.   This model prediction is shown in Fig.~\ref{fig:dilationstress} where $\delta$ is plotted vs. the predicted stress scale $\gamma/r(\delta)$.   A free parameter for the scale factor of 0.14 on the stress scale is used to fit the data.  The qualitative agreement in the model slope with the data in Fig.~\ref{fig:dilationstress} confirms that the confining stress scaling as $\gamma/r$ is a good estimate for the measured shear stress.   The fit coefficient within an order of magnitude of 1 confirms that the amount of dilation is on the right scale to provide the measured stress.    

The dotted line in Fig.~\ref{fig:dilationstress} corresponds to the dilation value where the contact line is expected to reach the 2nd layer of particles from the surface (see appendix for calculation).   At this point the confining stress should increase rapidly as more contacts are made with small curvature.  Because the calculation of confining stress from dilation is not monotonic around this region, the dilation is not single-valued function of confining stress.  The lower portion of the curve is expected to be unstable since more dilation would provide less of the stress required to confine the suspension to a smaller volume.  The agreement of the dilation measurements with the dotted line beyond the point where the calculation becomes multi-valued supports this interpretation.

As the dilation increases and the contact line recedes further into the interior, a lower limit for the value for the curvature must be reached as it is limited by the interstitial gap size. The corresponding limiting confining pressure has been measured in an analogous system in which a fluid interface was driven through a porous medium, in which case the required driving pressure went to $0.7\gamma/a$ in the limit of zero flow rate \citep{WSBK87}.  This confining stress sets the scale for the upper bound on the component of the shear stress due to capillary forces on the order of $\gamma/a$ in the limit of large $\delta$, although the exact value of the coefficient should depend on material specific properties such as contact angle and particle roughness.  The value of $\tau_{max}= 2$ Pa corresponding to the data in Fig.~\ref{fig:dilationstress} is significantly below the limiting confining stress regime, suggesting that dilation by a fraction of a particle width was enough to obtain a fully developed shear flow and the limiting confining stress is not necessarily reached in the shear thickening regime.


\subsection{Surface tension scaling}

\begin{figure}                                                
\centerline{\includegraphics[width=2.75in]{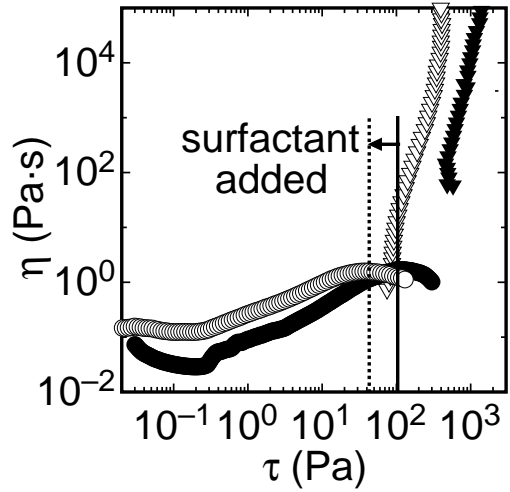}}
\caption{Viscosity curves for 100 $\mu$m glass spheres in liquids with different values of surface tension.  Solid symbols:  particles were suspended in water.  Open symbols: particles were suspended in water with surfactant (above the critical micelle concentration).  Triangles: $\phi=0.58>\phi_c$.  Circles: $\phi=0.56<\phi_c$.  Solid line: $\tau_{max}$ without surfactant.  Dotted line: $\tau_{max}$ with surfactant.  Both $\tau_{max}$ and the yield stress above $\phi_c$ decreased when the surface tension was reduced.  } 
\label{fig:stressmaxsurfacetension}                                        
\end{figure}

To confirm the role of capillary forces, we performed a set of rheological measurements in which we varied the surface tension of the liquid-air interface.  To vary this surface tension, we added surfactant to 100 $\mu$m glass spheres in water.  The surfactant used was Palmolive dish detergent, which was first mixed in water above the critical micelle concentration which reduces the surface tension with air by about a factor of 3 compared to pure water and air.  Viscosity curves are shown with and without surfactant and at different packing fractions in Fig.~\ref{fig:stressmaxsurfacetension}.  We first compare the viscosity curves for jammed suspensions at $\phi=0.58 >\phi_c$.  These viscosity curves correspond to yield stress fluids.  The value of the yield stress is reduced by a factor of 2.4 with the addition of the surfactant, about the same as the surface tension was reduced.

We next compare the viscosity curves at $\phi=0.56 < \phi_c$ in Fig.~\ref{fig:stressmaxsurfacetension}.  The increase in the viscosity at low shear rates can be attributed to the increase in the zero shear viscosity with the addition of the surfactant.  In terms of stress scales, there is a decrease in $\tau_{max}$ by a factor of $2.4$ when the surfactant is added, and no resolvable change in $\tau_{min}$.   The reduction in both $\tau_{max}$ and the yield stress $\tau_y$ above $\phi_c$ is comparable to the reduction in surface tension with the addition of surfactant, again consistent with a model in which these stresses scale with surface tension.

We note that in principle the addition of surfactant can change other relevant parameters.  The stress from surface tension on a boundary typically scales as $(\gamma/r)\cos\theta$ where $r$ is the radius of curvature and $\theta$ is the contact angle, where both $\gamma$ and $\theta$ can vary with the addition of surfactant.  The addition of a surfactant can reduce $\theta$, increasing the stress from surface tension.  However, we start with a liquid that wets glass pretty well, as this is a requirement to observe shear thickening \citep{BFOZMBDJ10}, so $\cos\theta\approx 1$ even before the addition of surfactant.  The addition of surfactant can also affect the value of $\tau_{min}$ in cases where the particle-liquid surface tension is dominant \citep{BFOZMBDJ10}, but for these 100 $\mu$m glass spheres in a wetting liquid the dominant force affecting the onset stress is gravity (Fig.~\ref{fig:sizestressgravity}).
 

\begin{figure}                                                
\centerline{\includegraphics[width=4in]{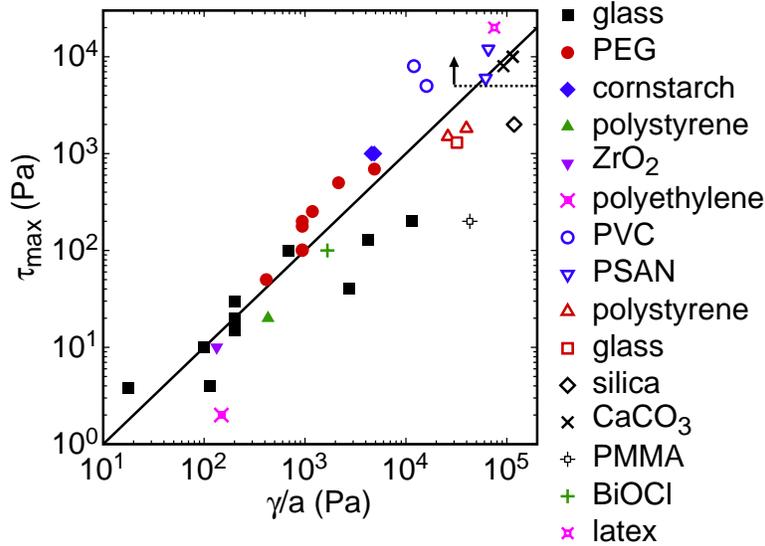}}
\caption{The stress at the upper bound of the shear thickening regime $\tau_{max}$ for a variety of suspensions plotted against the confining stress scale from surface tension $\gamma/a$.  Particle materials are listed in the key.  Solid symbols: measured by us.  Open symbols: polyvinyl chloride [PVC, circles \citep{Ho72}], polystyrene-acrylonite [PSAN, down-pointing triangles \citep{Ho72}], polystyrene [up-pointing triangles \citep{BBLS91}], glass [square \citep{BLS90}], silica [diamond \citep{BW96}], CaCO$_3$ [diagonal crosses \citep{EW05}], PMMA [crossed square \citep{KMMW09}], BiOCl [cross \citep{BBS02}], latex [diagonally crossed square \citep{LBS91}].  The solid line corresponds to a scaling $\tau_{max} = 0.1\gamma/a$.  Dotted line: lower bound on $\tau_{max}$ for measurements in which $\tau_{max}$ was not reached \citep{MW01a}, which often occurs in colloid measurements.
} 
\label{fig:stressmaxcollection}                                        
\end{figure}

The results in Fig.~\ref{fig:stressmaxsurfacetension} suggest that the upper stress scales $\tau_{max}$ and $\tau_j$ (the scale of the yield stress for $\phi>\phi_c$) scale with the surface tension at the liquid-air interface.  To more generally test the predicted stress scale $\tau_{max}\sim \gamma/a$ including the particle size scaling, we plot measured values of $\tau_{max}$ vs. $\gamma/a$ for the wide range of Discontinuous Shear Thickening suspensions we have studied in Fig.~\ref{fig:stressmaxcollection}.  Each point corresponds to a different suspension, with a wide range of different particle materials, shapes and sizes, and different liquids.  The plotted value of $\tau_{max}$ is an average over viscosity curves at several packing fractions.  We also included data from other papers in cases where $\tau_{max}$ was measured.  It is seen that for this wide variety of suspensions, covering four orders of magnitude, $\tau_{max}$ falls in a band that scales as the prediction $\gamma/a$ indicated by the solid line.  We note that for each Discontinuous Shear Thickening suspension we studied, the two stress scales $\tau_{max}$ and $\tau_j$ are always within an order-of-magnitude of each other, as was seen, for example, in Fig.~\ref{fig:stressmaxsurfacetension} and \citet{BJ09}, suggesting that $\tau_j$ also scales with $\gamma/a$.  In many measurements of colloids, the upper end of the shear thickening regime was not reached.  If there is an upper bound, it would have to be above the range measured.  This is especially a problem with colloids because the expected scale of $\tau_{max}$ for small particles exceeds the measuring range of many rheometers.   For example, our Anton Paar MCR 301 rheometer has an upper limit of 3800 Pa for the Couette cell or 65,000 Pa for the 25 mm diameter parallel plate.  This lower bound on $\tau_{max}$ based on the limited measuring range is illustrated as the dotted line in Fig.~\ref{fig:stressmaxsurfacetension}, using data from \citet{MW01a} as an example.

There is variation in the value of $\tau_{max}$ in the band shown in Fig.~\ref{fig:stressmaxcollection} by about an order of magnitude. There are many factors that could contribute to the precise value of the confining stress and the resulting shear stress.  For example, the normal stresses do not have to be exactly the same on each surface as would be the case for a pressure acting on a fluid.  Instead the stresses are related by a coefficient of order 1 \citep{Janssen1895, Sperl06}.  Since the confining stress can put a normal stress on the top plate via chains of particle contacts, then a component of the shear stress comes from friction, related to the normal stress by an effective coefficient of friction as seen in Fig.~\ref{fig:normalstress} which can depend on many factors including particle shape, roughness, and particle interactions.  The contact angle $\theta$ has been left out of the force equation since it is not known in many cases.  The dependence of dilation on the shear rate must also play a significant role, as seen in Fig.~\ref{fig:dilationstress}.  Geometric factors including particle shape and roughness also should play a role that has not yet been studied.  Considering all of these dimensionless factors of order 1 that can affect the shear stress which are not all known or easily measured, we will not go beyond using the order-of-magnitude stress scale of $\gamma/a$ as an estimate for $\tau_{max}$.  

In this section we proposed that Discontinuous Shear Thickening is due to a confining stress from surface tension in response to deformation of the liquid-air interface at the boundary from dilation.  We confirmed the stress increases with dilation, the effect of surface tension, and showed that $\tau_{max}$ scales with the confining stress scale $\gamma/a$ from surface tension with more than 30 suspensions that cover four decades of particle size.

\section{Solid boundaries}
\label{sec:wall}

In the previous section we showed that under boundary conditions such that particles penetrate the liquid-air interface, surface tension provides the confining stress that is responsible for Discontinuous Shear Thickening.  While a liquid-air interface at the boundary is typical for rheometer measurements, closed systems with solid walls are also of interest.  In this section we will use a solid-walled rheometer setup to investigate the role of the confining stresses in closed systems. 

\begin{figure}                                                
\centerline{\includegraphics[width=3.25in]{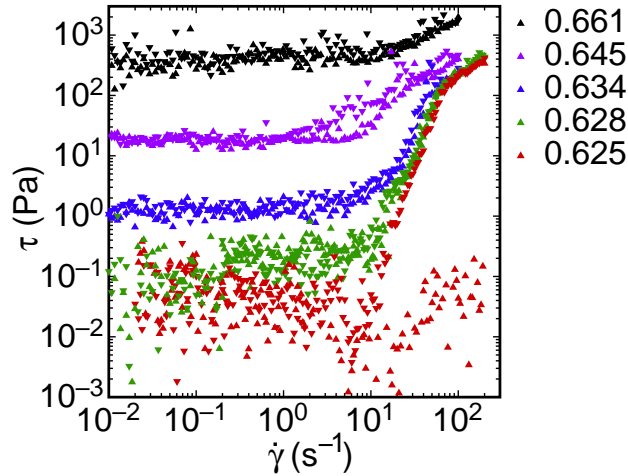}}
\caption{Stress vs. shear rate curves for 500 $\mu$m diameter glass spheres in a solid-walled rhoemeter with {\em no liquid}.  Packing fractions $\phi$ shown in the key; higher curves correspond to larger $\phi$.  Discontinuous Shear Thickening is still seen, confirming that viscous interactions are not necessary. 
} 
\label{fig:viscdry}                                        
\end{figure}

For measurements in a closed system, we used the parallel plate setup with solid walls shown in Fig.~\ref{fig:rheometer}b.  The hard walls confined large grains within the container volume without the need for the surface tension of the liquid.  Thus, we can also determine the role of the liquid by comparing measurements with and without liquid.  We first show stress vs.~shear-rate curves for shear rate controlled measurements of dry 500 $\mu$m glass spheres in Fig.~\ref{fig:viscdry}.  Without liquid, the packing fraction is determined by the container volume which can be varied with the gap size.  Thus, for a series of measurements with a fixed volume of particles, the gap height determines the packing fraction, with smaller gaps corresponding to higher packing fractions.  We give packing fraction values accurate to 3 decimal places relative to each other to compare curves in Fig.~\ref{fig:viscdry}, but absolute uncertainties on packing fractions are still around 0.01.

An important result from Fig.~\ref{fig:viscdry} is that the curves show Discontinuous Shear Thickening that is qualitatively similar to measurements of suspensions in standard rheology setups, despite the fact that there is no liquid.  Thus, the interstitial liquid or viscous stresses are not a necessary component for shear thickening when the grains are confined by other means.   

A large hysteresis loop can be seen for $\phi=0.625$ in Fig.~\ref{fig:viscdry}.  This is the threshold beyond which -- at larger gap sizes, corresponding to lower packing fractions -- not only the yield stress but also the measured shear stress was below the resolution limit suggesting contact between the plate and grains was lost.  This emphasizes that a key role of the liquid is simply to keep contact with the plates and transmit stress between the particles and the plate.

We repeated these measurements with water as a solvent filling the measurement volume and the surrounding volume so there was no liquid-air interface near any particles.  With water, contact between the suspension and plates could be maintained at larger gaps (lower packing fractions).  However, no significant difference was seen in the qualitative aspects of Discontinuous Shear Thickening or in the scale of $\tau_{max}$ with or without water.  Notably, the scale of $\tau_{max}$ with the solid wall is almost two orders of magnitude higher than for the same suspension in the parallel plate setup with a liquid-air interface (Fig.~\ref{fig:stressmaxcollection}).

An upper bound on the jamming transition can be identified by the point where the shear stress drops below the measurement resolution at $\phi=0.62$.  The jamming transition can be at a significantly higher packing fraction dry than with liquid because of the larger density difference between the particles and surrounding fluid \citep{OL90}.  With a solid wall, the yield stress did not plateau at high packing fractions like in the case of a liquid-air interface, but rather increased dramatically as the packing fraction was increased as seen in Fig.~\ref{fig:viscdry}.  This continued up to the maximum stress the rheometer can apply.  This can be expected if the confining stress comes from the stiffness of either the wall or the particles, in which case the confining stress increases as the solids are further compressed \citep{OSLN03}.  This is in contrast to the confining stress from surface tension which is limited by the capillary stress which reaches a plateau value as packing fraction is increased beyond jamming \citep{BJ09} due to the fact that the minimum radius of curvature is set by the particle size.

\begin{figure}
\centerline{\includegraphics[width=2.5in]{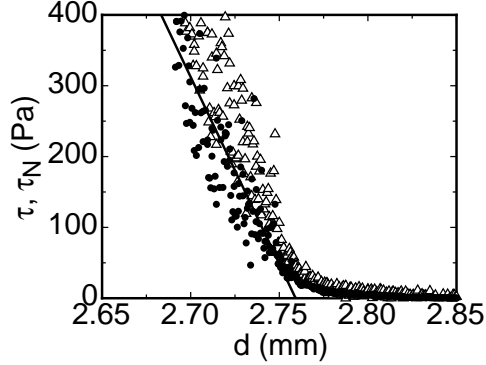}}
\caption{Shear stress $\tau$ (open triangles) and normal stress $\tau_N$ (solid circles)vs. gap size $d$ for a sample of 500 $\mu$m glass spheres with no liquid under slow compression with solid walls.    The sample is compressed at a rate of 0.25$\mu$m/s and sheared at a rate of 1 Hz (3mm/s).  The shear stress is close to the zero shear rate limit, so the measured $\tau$ is a good proxy for the yield stress.  The solid line is a linear fit used to obtain the per particle stiffness $k$ of the system of sheared grains and solid wall in series.
} 
\label{fig:compressiondry}                                        
\end{figure}

To connect $\tau_{max}$ to a confining stress for closed systems, we made measurements of the compressional stiffness of the tool and sample in series.  We observed that the rearrangement of particles under shear makes the suspensions much more compliant than under compression alone.  Thus we sheared the samples while measuring the compressional stiffness to better match the usual experimental conditions. The sample was slowly compressing at a fixed rate of 0.25 $\mu$m/s while also shearing at a fixed rate of 1 Hz (3 mm/s).  The shear rate was much faster than the compression rate so that the packing has time to rearrange as it is being compressed, but slow enough that the shear stress is still near the zero shear rate limit as seen in Fig.~\ref{fig:viscdry}.  The measured shear and normal stresses are shown in Fig.~\ref{fig:compressiondry}.   We note that the stiffness under shear is much less than the value obtained by compressing the sandpaper by itself, which is the weakest component of the wall.  Thus the presence of the grains has a significant effect on the effective stiffness, despite the fact that the material stiffness is much higher than that of the sandpaper.

In the case with a liquid-air interface, we found the confining stress to scale roughly as $-\delta\gamma/a^2$ due to linear compression of the boundary from dilation (Fig.~\ref{fig:dilationstress}).  If for the solid walls, the confining stress also comes from compression of the boundary, it should provide a restoring stress in response to dilation of $\delta\partial\tau/\partial d$.  To obtain an analog for the surface tension so that the confining stress scale can be written as $\delta k/a^2$, we define a stiffness per particle as $k= -a^2\partial\tau/\partial d$.  This differs from the usual definition of stiffness for an elastic material; rather than being proportional to wall surface area, it is normalized for a wall whose cross-sectional area is $a^2$, near that of a particle.  Since this normalization for the per-particle-stiffness makes it independent of the system size, it is more physically relevant in discussions of stress scales.  We can obtain $k$ from the slope in Fig.~\ref{fig:compressiondry}. 

\begin{figure}                                                
\centerline{\includegraphics[width=2.25in]{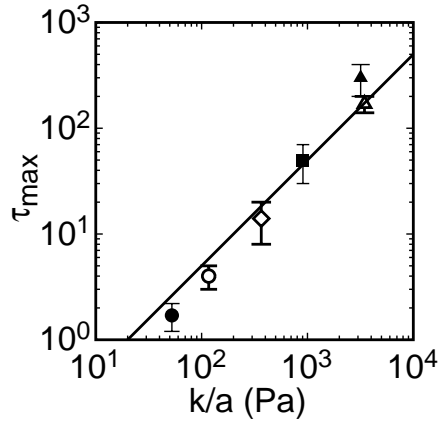}}
\caption{Maximum stress of the shear thickening regime $\tau_{max}$ vs. the confining stress scale $k/a$ due to the restoring force of the boundary with per-particle stiffness $k$.  Data are for 500 $\mu$m glass spheres under several different boundary conditions.   Solid triangle: hard wall rheometer setup, with particles suspended in water.  Open triangle: hard wall, no liquid.  Solid square:  hard wall with a soft foam rubber insert, with particles suspended in water.  Open circle: in a standard parallel plate setup with a liquid-air interface, where we use the surface tension to represent the per particle stiffness (i.e. $k=\gamma$).   Solid circle:  polyethylene in mineral oil from Fig.~\ref{fig:dilationstress} where we calculate $k=a^2(\partial\delta/\partial\tau)^{-1}$. The solid line has a slope of 1, corresponding to a stress response proportional to the restoring force of the boundary against a typical dilation of the sample by $\delta \approx 0.05a$. 
}
\label{fig:wallstress}                                        
\end{figure}

We plot values of $\tau_{max}$ vs. a confining stress scale equal to the boundary stiffness per particle over particle size $k/a$ in Fig.~\ref{fig:wallstress} each for the wet and dry 500 $\mu$m glass spheres.  We also measured a set of stress/shear-rate curves and compression curves for the wet glass spheres with a layer of soft foam rubber inserted between the top plate and sample as shown in Fig.~\ref{fig:rheometer}b, and the corresponding values of $\tau_{max}$ and $k/a$ are also plotted in Fig.~\ref{fig:wallstress}.  We also measured a set of data without sandpaper on the plate surfaces, which resulted in a much lower stiffness than with the sandpaper, despite having a harder surface, suggesting significant slip.   For comparison to the standard parallel plate measurements with a liquid-air interface, we plot $\tau_{max}$ vs. $\gamma/a$ for one such experiment with the same particles, using the surface tension $\gamma$ as a proxy for the stiffness per particle, which was also plotted in Fig.~\ref{fig:stressmaxcollection}.  As a test comparing to directly observed dilation, we plot $\tau_{max}$ vs. an effective stiffness per particle taken from Fig.~\ref{fig:dilationstress} as $k = a^2(\partial\delta/\partial\tau)^{-1}$ in the nearly linear regime.  All of these experiments are consistent with the relationship $\tau_{max} = 0.05 k/a$.  This scaling confirms that, for a wide range of boundary conditions including both liquid and solid boundaries, and even when there is a large amount of slip, the scale of $\tau_{max}$ is set by the confining stress which is proportional to the per particle stiffness of the boundary.  The similar values obtained when using $k/a$ or $\gamma/a$ as the effective stiffness confirms that the per-particle stiffness $k$ generalizes the role of the surface tension with a liquid-air interface to the case with a solid wall.  This means that the slopes of Figs.~\ref{fig:stressmaxcollection} and \ref{fig:wallstress} are related, although with the value of the coefficient relating $\tau_{max}$ and $k/a$ being suspension dependent.  The comparison with direct measurements of dilation from Fig.~\ref{fig:dilationstress} confirms that the confining stress could be written directly as $\tau_{conf} = \delta k/a^2$ for a linear elastic boundary. 

Since $k/a$ represents the restoring stress from the boundary for $\delta=a$, and the coupling coefficient between shear and normal stresses tends to be close to 1 (Fig.~\ref{fig:normalstress}), the slope of 0.05 between $\tau_{max}$ and $k/a$ suggests the restoring stress of the boundary is against a dilation of approximately $\delta \approx 0.05a$ at $\tau_{max}$.  This is the order of the expected dilation required to mobilize a single layer of shear, for example to allow spherical particles to escape out of the wells they can sit it at the interstices between three neighboring particles.  Similarly, dilation on the order of a percent of the sample thickness is typical of measurements of sheared granular packings \citep{Re1885, OL90}. 


In this section, we showed that for closed systems, the maximum stress in the shear thickening regime $\tau_{max}$ is determined by the wall stiffness.  In the experiments with a liquid-air interface, the stiffness of the boundary at the side determined the confining stress, while in the solid wall experiments, the stiffness of the top wall with the soft layer determined the confining stress.  In each case, the stiffness of the most compliant boundary determined the confining stress.  This is due to the fact that the stiffness of a collection of materials in series is determined by the most compliant material.  Whether the most compliant boundary is on the side or the top does not matter, since the compressional relation between shear and normal stresses suggests that the stresses are similar on all of the boundaries \citep{Janssen1895}. 


\section{Discussion}
\label{sec:discussion}

\subsection{Non-local constitutive relation}
\label{sec:constitutivelaw}

In this subsection, we combine the results of the previous sections to write the full constitutive relation for the mechanism we have found for Discontinuous Shear Thickening suspensions.  In Sec.~\ref{sec:shearprofile} we showed a constitutive relation for the shear stress accounting for viscous and  gravitational forces.  The shear profile measurements implied that the dramatic increase in stress associated with Discontinuous Shear Thickening could not be attributed to a local shear rate dependence from viscous stresses, whose contribution appeared to be relatively constant in the shear thickening regime.  Instead, we showed in Sec.~\ref{sec:normalstress} that the shear stress was coupled to the normal stress, with a proportionality given by an effective friction coefficient $\mu$.  In turn, the normal stress was shown to come from the restoring force of the boundary in response to dilation (Secs.~\ref{sec:surfacetension}, \ref{sec:wall}).  We can now fully express the constitutive relation for local shear stress $\tau_l$ based on Eq.~\ref{eqn:constitutivelaw} with the uniform term $\tau_c$ explicitly separated into the confining stress contribution as well as a constant contribution from other interparticle interactions $\tau_{int}$: 

\be
\tau_l = \eta_{\nu}(\phi)\dot\gamma_l f(Re) + \tau_g h/d + \mu \tau_{conf}(\delta)+\tau_{int} \ .
\label{eqn:constitutivelaw2}
\ee

\noindent  The stress terms represent, from left to right,: viscous and inertial hydrodynamics, gravity, confinement, and particle attractions.  The confinement term simplifies to $\tau_{conf}(\delta) = \delta k/a^2$ for a linear elastic boundary with per-particle stiffness $k$ in response to dilation $\delta$ (Fig.~\ref{fig:wallstress}).  The hydrodynamic term varies with Reynolds number such that $f(Re) =1$ in the viscous regime where $Re < Re_c \sim 100$ and $f(Re) \propto  Re/Re_c$ in the fully inertial regime where $Re \gg Re_c \sim 10^2$ (Sec.~\ref{sec:inertial}).  

While this constitutive relation is approximate, its main use is that it allows for simple estimates of the stress scales $\tau_{min}$, $\tau_{max}$, and $\tau_j$ in different regimes, and their comparison can easily determine whether a system will shear thin or thicken.   This includes, for example, the requirement that the increase in stress in the confinement term in response to dilation must be large compared to any yield stress from the attraction or gravity terms \citep{BFOZMBDJ10}.  It also delineates the packing fraction regimes that separate inertial and Discontinuous Shear Thickening as seen in Fig.~\ref{fig:reynolds}.  The difference between the boundary conditions of fixed gap and fixed normal force as in Fig.~\ref{fig:normalstress} can be characterized by whether the confinement term is fixed by the boundary or varies in response to dilation.

The local constitutive relation can also simply explain observations under different measuring geometries, for example in suspensions where gravity is dominant.  In a parallel plate geometry, we found settling suspensions to exhibit an apparently Newtonian scaling behavior below $\tau_{min}$ because the plate measures the shear stress at a depth $h=0$.  In a Couette geometry where the moving wall is on the side, any density mismatch was found to result in a yield stress scaling as $\tau_g \sim \Delta\rho g H$ \citep{FBOB09},  since the average depth at the moving wall is $\langle h\rangle = H/2$.



  
 It is interesting to note that the direct shear-rate dependence in Eq.~\ref{eqn:constitutivelaw2} is inherently shear thinning in the low-$Re$ regime where Discontinuous Shear Thickening is observed.   Similarly, we found that the local viscosity in the bulk obtained from shear profile measurements corresponded to shear thinning (Figs.~\ref{fig:shearprofileZrO},\ref{fig:shearprofilepolyethylene}), even for a global mechanical response that corresponded to shear thickening.  The explanation for this apparent contradiction is that most of the shear stress is due to frictional contacts and normal stress in response to frustrated dilation, which only indirectly depends on shear rate.  This response is non-local in the sense that the response depends on the global dilation and boundary conditions.  Dilation can be treated as a global rather than local parameter as long as the boundary is linearly elastic, which is a good approximation for most solid boundaries and for surface tension where $k=\gamma$ (Fig.~\ref{fig:dilationstress}), so the average normal stress from the boundary is proportional to the average strain from dilation, regardless of inhomogeneities.  Local variations only need to be accounted for if the boundary stiffness is non-linear.   
  
 It is notable that the boundary conditions can have an uncommonly large effect on the normal stress term because it is due to a confining stress.  From a hydrodynamic point of view, the large significance of the boundary conditions and difference between local and global results is unusual, as it is more typical for the boundary conditions to play a smaller role, requiring only perturbative corrections to translate between the local and global rheology.  One of the surprising consequences of this is that characterizing rheology based solely on local, shear-rate dependent constitutive laws or local viscosities in the bulk would miss the dramatic phenomenon associated with Discontinuous Shear Thickening.
 
We cannot yet fully solve the constitutive relation in Eq.~\ref{eqn:constitutivelaw2} because dilation in suspensions is not yet well-characterized in the Discontinuous Shear Thickening regime of concentrated suspensions.  Specifically, the slope of viscosity curves in the shear thickening regime should depend on how dilation couples with shear rate, stresses, and packing fraction.  Dilation has been characterized in suspensions that are dominated by viscous interactions and are not so confined as to frustrate dilation \citep{PK95, SB02, DGMYM09}.  On the other hand, dilation in dry grains does not occur until significantly higher packing fractions, at least in the quasi-static limit \citep{OL90,JSSSSA08,KS09}.  We speculate that the hydrocluster models may be able to bridge this gap by describing the transition from viscous flow to frustrated dilation with frictional contacts \citep{OM00}.   Perhaps if hydrodynamically-induced particle clusters  \citep{BB85} were to become large enough to span the system become jammed against the boundaries, they could lead to the frictional contacts and dilation.

 
The shear profile measurements of Sec.~\ref{sec:shearprofile} suggests a possible connection between shear banding and dilation.  Fig.~\ref{fig:dilationstress} shows a single scaling for the dilation with shear stress through the transition at $\tau_{max}$.  Since the stress/shear-rate curve corresponds to Discontinuous Shear thickening, this implies the dilation $\delta$ increases rapidly with shear rate in the shear thickening regime, then increases less rapidly above $\tau_{max}$.  Suggestively, this rapid increase of dilation with shear rate corresponds to the regime where the shear band is widening (Fig.~\ref{fig:shearprofileZrO}), and the dilation increases less rapidly with shear rate once the shear band stops widening.  We cannot say for sure that these observations are connected, but it may be that the rapid increase in dilation is required to involve more layers of particles in the shear flow.  
 
 

\subsection{Dominant stress scales for the onset}
\label{sec:stressscales}

In this subsection we discuss the different scaling laws for the onset stress $\tau_{min}$ found in different parameter regimes, including suspensions dominated by gravity and colloids dominated by Brownian motion and electrostatic forces.  We suggest that different scalings for $\tau_{min}$ can be understood as the dominant stress scales of particle interactions in each system, and these observations suggest a more general requirement for the onset of shear thickening.


We first compare the different scaling laws for the onset stress $\tau_{min}$ to identify common features of the different regimes.  In Sec.~\ref{sec:onsetstress}, we found that for suspensions of particles large enough to settle the scale of $\tau_{min}$ is set by gravity such that the onset of shear thickening required enough shear stress to lift the weight of the top layer of particles to initiate shear and dilation (Fig.~\ref{fig:sizestressgravity}).  In a previous work we considered the effect of induced particle attractions in response to external electric and magnetic fields \citep{BFOZMBDJ10}.  In each case the attractions resulted in a yield stress. The scale of $\tau_{min}$ was set by the shear stress required to overcome roughly the two-particle attractive force (per cross-sectional area of a particle) to shear them apart.  In both the gravity and attraction regimes the shear stress must simply exceed all local stress barriers from any source that are responsible for preventing relative shear between particles.  While this generalization is sufficient for frictional systems, hydrodynamic systems can exhibit shear at stresses significantly below the onset of shear thickening.  One such regime is colloids stabilized by an electrostatic zeta potential $\zeta$.  \citet{MW01a} measured $\tau_{min}$ for such particles from 80 to 700 nm in diameter.  A power law fit to their measurements for the onset of shear thickening gave $\tau_{min} \propto a^{-2.11\pm0.16}$.  An electrostatic calculation of the two-particle repulsive force over the cross-sectional area of spherical particles gives a stress scale of $16\epsilon\zeta^2/a^2$ for a liquid permittivity $\epsilon$.  This has both the same scaling and magnitude within about a factor of 2 of their data, consistent with the idea that the scaling for the onset of shear thickening is determined by the interaction stress scale.  The same scaling was argued for by \citet{Ho98}, although with a somewhat different mechanism in mind.  In another regime dominated by Brownian motion, the onset of shear thickening has been found to correspond to an onset stress $\tau_{min} = 50kT/3\pi a^3$ \citep{MW01b, GZ04}.  This scale is the osmotic pressure, which is effectively the pressure which neighboring particles interact with.  The common feature In each of these scaling regimes is that  the onset of shear thickening corresponds to the dominant stress scale at which neighboring particles interact, whether they interact by induced dipoles, gravity, zeta potential, or osmotic pressure,  in each case.  With particular relevance to mechanisms for shear thickening, these stress barriers prevent {\em compressive} shear of particles into each other, and once these stress barriers are overcome, compressive interactions are possible which could lead to dilation against a confining boundary to produce Discontinuous Shear Thickening.  


It is notable that the scaling laws for the onset stress $\tau_{min}$ have been found to work both for Discontinuous and Continuous Shear Thickening \citep{MW01b,GZ04}, even though in some cases the mechanisms may be different.  We suggest the generality of the onset law arises because the scalings for $\tau_{min}$ come from interaction scales that are not directly responsible for shear thickening, rather they are preventing it.  This is why different proposed models have predicted the same scaling law for $\tau_{min}$ in the zeta-potential dominated regime, including the hydrocluster model \citep{MW01a}, the order-disorder transition model \citep{Ho98}, and the dilational model, even though the proposed mechanisms for shear thickening are very different.  The hydrocluster model which describes Continuous Shear Thickening also requires compressional flow between particles to occur, and thus the same stress scales would prevent both mechanisms.  Thus, the requirement for shear thickening that the shear stress exceed the various stress scales of particle interactions that prevent compressive shear seems to be valid for both Continuous and Discontinuous Shear Thickening, in both suspensions and colloids, and whether the particle interactions are frictional or hydrodynamic in nature, as long as the mechanism for shear thickening requires compressional interactions. 

Many of these same stress scales that determine the onset of shear thickening can also determine the yield stress \citep{BFOZMBDJ10}, with the exception of osmotic pressure.  Just above the yield stress, the confining stress can start to grow as shear causes dilation, but there must be at least a small shear thinning regime before the confining stress becomes dominant and shear thickening is seen \citep{GZ04, BFOZMBDJ10}.  Indeed, \citet{MW58} found the onset of dilatancy at slightly lower shear rates than the onset of shear thickening.  In other words, the onset of shear thickening corresponds to a transition in the dominance of different stress and not necessarily where the mechanism for shear thickening first appears.  

While we have described a mechanism for Discontinuous Shear Thickening that is based on generic phenomena such as dilation, not all suspensions and colloids exhibit this behavior at high packing fractions.  This can be explained by the relative importance of different stress scales.  If any other particle interaction scales exceed the confining stress from surface tension, we would expect shear thinning mechanisms to be dominant over shear thickening \citep{BFOZMBDJ10}.  Some of the interactions that we did not explicitly mention above include hydrogen bonding \citep{RWK00}, depletion \citep{GZ04}, or a particle-liquid surface tension \citep{Ba89, BFOZMBDJ10}.  One reason that many dense suspensions and colloids do not exhibit shear thickening is that many of these systems fall into regimes where other stress scales exceed the confining stress scale so they do not have any observable shear thickening regime.  


Another example of shear thinning stresses hiding shear thickening can be seen in Fig.~\ref{fig:sizestressgravity} for settling suspensions.  The maximum particle size at which shear thickening was found was about 1000 $\mu$m.  At such a large scale, the weight of the particles contributes to a large frictional stress $\tau_{min} \sim  \Delta\rho g a$, which increases with particle size.  This can overwhelm the scale of the confining stress from surface tension $\tau_{max} \sim \gamma/a$, which decreases with particle size.  The balance between these two stress scales occurs at a particle size $a \sim \sqrt{\gamma/(\Delta\rho g)} \sim 1000$ $\mu$m, above which the shear thinning stresses dominate, in agreement with the  maximum size particle found to shear thicken.
This transition scale could be interpreted as a particle capillary length scale which differs from the usual capillary length in two ways.   First, this particle capillary length depends on the density difference rather than just a liquid density, so density matching could allow for shear thickening of larger particles.  Second, this particle capillary length depends on particle size rather than system size.  This means surface tension effects can be seen in suspensions on much larger scales than would be expected based on the usual capillary length.  A similar result was found by the work of \citet{LNS02} on free-surface flows of dense suspensions, in which they found effective stresses from surface tension scaling as $\gamma/a$.

\subsection{Connection to confining stresses in other systems}
\label{sec:confiningstress}

In this subsection, we discuss similarities of the role of the confining stress to other systems, including colloids, soil mechanics, and jammed systems.  Similar to the case for $\tau_{min}$, we also suggest that different scalings for $\tau_{max}$ can be understood in a more general framework. 

We found the upper end of the shear thickening regime $\tau_{max}$ to be set by the maximum confining stress which comes from the restoring force when grains dilate against a boundary, either from surface tension when there is a liquid-air interface (Figs.~\ref{fig:stressmaxsurfacetension}, \ref{fig:stressmaxcollection}) or by the stiffness of the wall when all boundaries are solid (Fig.~\ref{fig:wallstress}).   More generally, this role of the confining stress requires only compressional interactions between particles that lead to dilation, whether the forces are transmitted via frictional interactions as we showed in Sec.~\ref{sec:normalstress}, or through lubrication interactions as suggested by hydrodynamic models.  The importance of the confining stress can also be independent of what mechanism initiates shear thickening as long as the particle interactions are compressive.

In the discussions of the confining stress so far, the most compliant boundary set the response.  This is because the stiffness of a system of several elements with very different stiffnesses in series will generally be determined by the most compliant element in the series.  In some systems, the particles could be the most compliant element.  This regime would be relevant when all of the boundaries are hard in comparison to the particles, and it has been proposed such a regime may be reached for small colloidal particles where the confining stress from surface tension is larger \citep{WB09}.  If there is a lubrication layer of liquid between gaps, the maximum confining stress would be coupled to the viscosity because the particle compression depends on the stress in the lubrication layer.  For solid contacts between elastic spheres, the confining stress would be limited by a scale of $(\delta/R)^{3/2}E_p$ where $E_p$ is the compressional modulus of the particles and $\delta/R$ corresponds to the compressional strain on the sample. The $3/2$ power comes from the contact between two spherical surfaces as opposed to the power of 1 for flat surfaces.  For the hard particles we used with $E_p \sim 10^{10}$ and $\delta\approx 10^{-2}$, this scale is of order $10^7$ Pa, which is much stiffer than the liquid-air interface.  Stresses up to about $10^7$ Pa have been observed in the shear thickening regime for silica particles in compressional flows with solid walls \citep{LLWG10}, which is consistent with the idea that much larger confining stresses can be reached for hard walls and particle stiffness becomes the limiting factor.  On the other hand, if the particles are extremely soft, the confining stress could be below $\tau_{min}$.  This seems to be the case in a suspension of soft gel particles with modulus on the order $10^4$ Pa in which only shear thinning was observed instead of shear thickening as the jamming transition was approached \citep{NVABZYGD10}.  So far, it is not certain whether or not Discontinuous Shear Thickening is controlled by a confining stress in the colloid regime.  Despite the comparison between measurements of $\tau_{max}$ and the confining stress from surface tension in Fig.~\ref{fig:stressmaxcollection} and the arguments for a particle-stiffness limited confining stress, there is still no direct experimental evidence with a systematic control of boundary or particle stiffnesses to conclusively determine whether Discontinuous Shear Thickening in colloids is limited by a confining stress. 



While simulations of suspension have successfully modeled Continuous Shear Thickening effects, so far most have failed to produce the large, steady-state stress increases associated with Discontinuous Shear Thickening \citep{BB85, MVB96, FMB97,  BBV02, MB04a, MB04b, GCNC08}.  These simulations have included viscous interactions as well as various interparticle interactions.  Most have focused on bulk behavior, usually using periodic boundary conditions such as Lees-Edwards to avoid dealing with boundary effects.  Now that we have recognized that the boundary conditions and especially the confining stress are important, it seems likely that many of these simulations did not find Discontinuous Shear Thickening because of their treatment of the boundary conditions.  The one simulation we are aware of that has produced Discontinuous Shear Thickening  was a molecular dynamics simulation of two-dimensional granular shear flow with frictional contacts between particles but no liquid or viscous interactions \citep{OH10}.  Besides a steep $\tau(\dot\gamma)$ at packing fractions just below the jamming transition, the scale of the normal and shear stresses was found to be set by the particle modulus, which in that simulation was the only scale that could limit a confining stress.  This may be a minimal model for Discontinuous Shear Thickening in two dimensions since it includes particle-particle contacts with a restoring force,  but leaves out the liquid.

While we have described a mechanism for shear thickening due to confining stresses in shear flows, the same principle could apply to extensional and compressional flows because stresses tend to be easily redistributed in different directions when the shear and normal stresses are proportional.  Visible dilation at the surface was seen to correspond to shear thickening in an extensional flow\citep{SBCB10}.  The maximum stress in the shear thickening regime $\tau_{max}$ in extensional flows \citep{CAR09, WCR10} has been found to be about an order of magnitude higher than $\tau_{max}$ based on shear measurements, but this could still be consistent with a scale of $\gamma/a$.  While these isolated result are promising, our model has not yet been extensively tested in extensional flows. 



Systems where the mechanics are determined by a confining stress are already well-known, for example, in granular systems, especially soil mechanics \citep{LW69}.  One lesson to take away from soil mechanics is that even though the global response of the system is set by the boundary conditions, the scale of the stress response is not dependent on sample size or shape. This is because forces will transmit throughout the bulk across particle contacts, and forces must balance across the system, regardless of how far across the bulk is.  This makes stress the appropriate size-independent force scale as in other continuum systems.  In Discontinuous Shear Thickening suspensions, the same qualitative behavior has been seen from as few as 2 particle layers \citep{BZFMBDJ10} to tens of thousands of layers \citep{MW01a}.  Quantitatively, the significance of the surface area to volume ratio can be checked by varying the gap size in a parallel plate geometry for a fixed volume of sample.  For such measurements at constant shear rate, we found that the percentage change in stress was $(0.20\pm0.22)\%$ over a range where the surface area changed by 17\% \citep{BZFMBDJ10}.   This is consistent with a shear stress independent of surface area and inconsistent with a stress proportional to surface area.  Furthermore, similar results have been found when comparing Couette cell measurements with parallel plate measurements \citep{FHBOB08}.  Together, these observations support the argument that the relevant stress scale for shear thickening is not dependent on the system size or shape.

It has been suggested that Discontinuous Shear Thickening is a form of jamming \citep{FMB97, CWBC98, HL05, MW01a, FHBOB08}.  Visible shear in and above the shear thickening regime shows that shear thickening is not jamming in the sense of being associated with a yield stress or static structures (Figs.~\ref{fig:shearprofileZrO}, \ref{fig:shearprofilepolyethylene}, \ref{fig:dilationtangentview}).  Below the onset of shear thickening, the particles may be settled or stuck together by attractions; in either case this corresponds to a locally static structure, and in many cases the system is jammed with a yield stress below the shear thickening regime.  The shear thickening regime can be where more of the particles become involved in shear,  i.e.~becoming unjammed in the above sense, and it is the resulting dilation against the boundaries under shear which is responsible for Discontinuous Shear Thickening.  

One connection between Discontinuous Shear Thickening and jamming comes from the observation that they are controlled by the same critical packing fraction $\phi_c$ \citep{BJ09}.  The shear rate at the onset of shear thickening goes to zero in the limit of $\phi_c$, suggesting the limiting case of shear thickening corresponds to a yield stress, i.e. a jammed state, which is also what is found on the other side of $\phi_c$ \citep{BJ09}.  It is suggestive that this is also near the critical point for the onset of dilation in dry granular systems \citep{OL90}, and the increase in dilation with shear rate must become steeper as this critical point is approached for our model to fit the data.  Whether this feature is relevant for understanding dilation at non-zero shear rates in dense suspensions is unclear at this point.  The other connection between Discontinuous Shear Thickening and jamming, which was hinted at by \citet{HFC03,HCFS05} and \citet{MB04b},  comes from the stress response, whose scales $\tau_{max}$ and $\tau_j$, respectively, are set by a confining stress due to the penetration of particles through the liquid-air interface in each case and which is transmitted through the system via force chains.  This occurs with dilation under shear for Discontinuous Shear Thickening (Figs.~\ref{fig:dilationtangentview}, \ref{fig:dilationstress}), and at rest for jammed suspensions \citep{BZFMBDJ11}.  In terms of the stress response, Discontinuous Shear Thickening could be considered to be a dynamic extension of jamming, but so far there is not yet a formalism for describing this.

 \subsection{Summary of the mechanism for Discontinuous Shear Thickening}
 \label{sec:mechanism}
 
The requirements for Discontinuous Shear Thickening under the mechanism we proposed, not necessarily limited to suspensions, can be summarized as:
\begin{enumerate}
\item \underline{Dilation}: The particle packing must attempt to dilate with increasing shear rate.
\item \underline{Frustration}:  There must be geometric constraints that at least partially frustrate dilation.  These constraints can come from surface tension at a liquid-air interface, the walls surrounding the system, or the particle stiffness, for example.  This frustration usually requires a high packing fraction, close to the jamming transition.
\item \underline{Confining stress}: These constraints must provide a confining stress such that the most compliant constraint produces a restoring force that increases with dilational strain.
\item \underline{Dominance}:  This confining stress must significantly exceed all stresses that prevent shear between grains and dilation, such as interparticle interactions or gravity.  Otherwise, there is not enough stress increase from dilation to result in a positive slope on a viscosity curve and the global rheology may be shear thinning instead.
\end{enumerate}
 
Here we compare this mechanism to others that have been proposed for colloids and suspensions.  \citet{Ho82} argued that ``[Discontinuous Shear Thickening] will occur in concentrated suspensions whenever the particles can segregate into layers parallel to planes of constant shear, but are constrained from free rotation at levels below some critical level of stress."  Our model is in agreement with that of \citet{Ho82} in the sense that the onset stress is determined by a point where the shear stress is large enough to shear particles in such a way to cause dilation.  \citet{Ho82} argued for the same onset scaling law with the stress scale corresponding to the interaction between zeta potentials of neighboring particles.   Our proposed scaling for the confining stress $\tau_{max}$ also gives the scale measured by \citet{Ho72}, whose data is included in Fig.~\ref{fig:stressmaxcollection}.  While Hoffman observed an order-disorder transition along with Discontinuous Shear Thickening, it was shown that such a transition is not necessary \citep{MW02, EW05, ENW06}, which suggests the order-disorder transition is not the key to explaining the stresses.  

Many other papers have supported a hydrocluster mechanism \citep{BB85, MW01a, SW05} based on the ability of those models to predict some of the scalings for the onset $\tau_{min}$.  While those models have been successful at predicting weaker, continuous shear thickening, we suggest that for the much higher stress scales found in Discontinuous Shear Thickening to be reached, an additional mechanism is likely required to introduce another stress scale; for example, compression against the boundary introduces a confining stress scale from the boundary stiffness.  The hydrocluster model has been successful at describing the onset stress $\tau_{min}$ for Discontinuous Shear Thickening in some cases, but we have argued that the scalings for both $\tau_{min}$ and $\tau_{max}$ predicted by different models for shear thickening can be more simply understood in terms of dominant stress scales, which are not dependent on whether stress transfer is through frictional or hydrodynamic interactions, or what mechanism initiates shear thickening.  In this discussion, we have suggested that the dilational model based on suspensions could also apply to the colloid regime.  While we have been able to compare our model to known scalings for $\tau_{min}$ and to measurements of $\tau_{max}$ (Fig.~\ref{fig:stressmaxcollection}), more data is needed to directly test the dilational model in the colloid regime.


 
 

\section{Conclusions}

In this paper we proposed and experimentally validated a model for Discontinuous Shear Thickening in suspensions, which we briefly summarize here.  The onset of shear thickening  $\tau_{min}$ occurs when the grain packing starts to dilate (Figs.~\ref{fig:dilationimages}, \ref{fig:dilationtangentview}), which requires that the applied shear stress overcome any interparticle stresses preventing compressive shear between particles (Figs.~\ref{fig:sizestressgravity}, \ref{fig:shearprofileZrO}, \ref{fig:shearprofilepolyethylene}).    Dilation in turn causes the particles to push against the boundary, typically the liquid-air interface for suspensions open to the air, which pushes back with a restoring force to produce a confining stress on the suspension (Figs.~\ref{fig:dilationstress}, \ref{fig:stressmaxsurfacetension}, \ref{fig:stressmaxcollection}).  The resulting normal stresses are transmitted through the packing via frictional interactions (Figs.~\ref{fig:profilequadfit}, \ref{fig:normalstress},\ref{fig:viscnormalforcecontrol}), resulting in a rapid increase in shear stress with shear rate corresponding to Discontinuous Shear Thickening.  We generalized this shear thickening mechanism to other sources of a confining stress by showing that, when instead grains are confined by solid walls and have no liquid-air interface, $\tau_{max}$ is set by the stiffness of the most compliant boundary (Figs.~\ref{fig:viscdry}, \ref{fig:compressiondry}, \ref{fig:wallstress}).  

These results allowed us to suggest a generalization of the scaling laws for the stresses that bound the shear thickening regime to systems in other parameter regimes: the stress required for compressional shear between particles for $\tau_{min}$ (Sec.~\ref{sec:stressscales}), and the confining stress for $\tau_{max}$  (Sec.~\ref{sec:confiningstress}).    We find that the experimental results can be described by a non-local constitutive relation where the stress does not come directly from the local shear rate, but where Discontinuous Shear Thickening comes out of a frictional term from the confining stress at the boundary which depends on the global dilation (Sec.~\ref{sec:constitutivelaw}).


\section{Acknowledgements}

We thank S. Nagel, T. Witten, and W. Zhang for thoughtful discussions.  We thank Marc Miskin for designing the optical setup used in the shear profile videos.  We thank Franco Tapia Uribe for taking preliminary shear profile measurements.  We thank Trevor Martin for machining the solid wall setup for the rheometer.  This work was supported by DARPA through Army grant W911NF-08-1-0209 and by the NSF MRSEC program under DMR-0820054.

\section{Appendix:  calculation of confining stress at liquid-air interface from dilation}

\begin{figure}
\centerline{\includegraphics[width=1.2in]{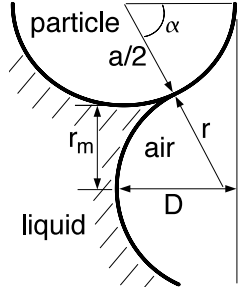}}
\caption{Dimensions used for calculation of relationship between dilation $\delta$ and confining stress due to surface tension at a liquid-air interface with radius of curvature $r$. }
\label{fig:dilationcalc}                                        
\end{figure}

To connect the confining stress at the liquid-air interface to the measured dilation $\delta$, a model is needed for the interface geometry.  For simplicity we will assume spherical particles with a contact angle $\theta=0$ at the liquid-solid-air contact line since to obtain shear thickening the liquid must wet the particles.  For this contact angle the interface geometry is equivalent to a sphere of air of radius $r$ in contact with particles at the surface.   We will calculate the geometry for a characteristic radius of curvature as if it is the same at each interstice between particles, and use these mean single-particle calculations as an estimate for the surface as a whole, ignoring variations in the surface curvature.  The point of contact is defined by an angle $\alpha$ relative to horizontal as shown in Fig.~\ref{fig:dilationcalc}.  The packing fraction of particles on the two dimensional surface will be represented by $\phi_{2D}$.  As an estimate we will use the value $\phi_{2D}=0.84$ which corresponds to random close packing in two dimensions \citep{OLLN02}.  This geometry gives enough constraints to relate the radius of curvature $r$ to the contact angle.  The dilation $\delta$ can be connected to this geometry using conservation of volume.  The confining stress scale from surface tension can then be calculated as proportional to $\gamma/r$ as a function of $\delta$.

From the geometry in Fig.~\ref{fig:dilationcalc}, the vertical components of the dimensions can be used to relate the radius of curvature to the contact angle

\be
r\sin\alpha = \frac{a}{2}(1-\sin\alpha) + r_m \ .
\label{eqn:curvalpha}
\ee

\noindent The minimum radius of curvature $r_m$ comes from the minimum radius of the insterstitial gap
between particles.  At an interstice between 3 particles in contact, this is $r_m/a = \sqrt{1/3}-1/2 \approx 0.26$ in three dimensions, while in two dimensions $r_m$ would be zero.  The mean confining stress is modeled as 

\be
\tau_{conf} \sim \frac{\gamma}{r} (1-\phi_{2D}\sin^2\alpha)
\label{eqn:tauconf}
\ee

\noindent where $1-\phi_{2D}\sin^2\alpha$  is the fractional cross-sectional area around the outer edge covered by the liquid-air interface.

Next, we relate the dilation $\delta$ to $\alpha$.  Conservation of volume requires that the dilation match the enclosed volume of air per particle $\Delta V$ up to the maximum penetration of the particle when using the initial condition that the surface is flat ($r=\infty$) when $\delta=0$:

\be
\delta = \frac{4\phi_{2D} \Delta V(r,\alpha)}{\pi a^2} \ .
\label{eqn:delta}
\ee

\noindent The volume $\Delta V$ can be calculated as 

\be
\Delta V = 2V_{cap} + V_i
\label{eqn:deltav}
\ee

\noindent where $V_{cap}$ is the volume of a spherical cap interior to the point of contact (of which there are 2 per particle) 

\be
V_{cap} = \frac{\pi r^3}{3}(1-\cos\alpha)^2(2+\cos\alpha)
\label{eqn:vcap}
\ee

\noindent and $V_{i}$ comes from integrating the fluid volume in the mean surface normal direction from the furthest point of penetration of the particle up to the point of contact:

\be
V_{i} =\int_0^{\alpha} \frac{\pi a^3}{4}\sin\alpha' \left[\frac{1}{\phi_{2D}} - \sin^2\alpha'\right] d\alpha'  = \frac{\pi a^3}{4}\left[\frac{1-\cos\alpha}{\phi_{2D}} - \frac{\cos(3\alpha) - 9\cos\alpha+8}{12}\right] \ .
\label{eqn:vint}
\ee


\noindent  With Eqns.~\ref{eqn:curvalpha}, \ref{eqn:tauconf}, and \ref{eqn:delta} relating $\tau_{conf}$, $\delta$, $r$, and $\alpha$, they can be solved numerically to obtain a relationship between $\delta$ and $\tau_{conf}$, with the result shown in Fig.~\ref{fig:dilationstress}.

These equations may be valid until the point where the contact line reaches the 2nd layer of particles from the surface, beyond which extra contact lines are made.  This starts when the maximum penetration, defined in Fig.~\ref{fig:dilationcalc} as $D=(a/2+r)(1-\cos\alpha)$, reaches $\sqrt{3}a/2$ which corresponds to the layer width for a hexagonal packing.  This limit is shown as the dashed line in Fig.~\ref{fig:dilationstress}. 

\end{document}